\documentstyle[12pt,dina4,amsfont12]{article}

\input psfig.sty

\renewcommand{\theequation}{\arabic{section}.\arabic{equation}}
 \newcommand{\bx}{{\bf x} }

\begin{document}

\title{Convergence rate of dimension reduction
in Bose-Einstein condensates }
\author{\small Weizhu  Bao$^1$, Yunyi Ge$^1$, Dieter Jaksch$^2$,
Peter A.~Markowich$^3$ and Rada M.~Weish\"{a}upl$^3$}

\date{}
\maketitle

{\small  $^1$ Department of Mathematics and Center for
Computational Science \& Engineering, National University of
Singapore, Singapore 117543. Email:
{\it bao@math.nus.edu.sg} (W.B.) and  {\it yunyi.ge@nus.edu.sg} (Y.G.)\\

$^2$Clarendon Laboratory, Department of Physics, University of
Oxford, Oxford, United Kingdom. Email: 
{\it Dieter.Jaksch@physics.ox.ac.uk.}\\

$^3$Instit\"{u}t f\"{u}r Mathematik, Universit\"{a}t  Wien,
Nordbergstr. 15, 1090 Vienna, Austria. Email: {\it
peter.markowich@univie.ac.at} (P.A.M.) and {\it
rada.weishaeupl@univie.ac.at} (R.M.W.)

}


\begin{abstract}

  In this paper, we study
 dimension reduction of the three-dimensional (3D)  Gross-Pitaevskii equation
 (GPE) modelling Bose-Einstein condensation under different limiting
 interaction and trapping frequencies parameter regimes.
 Convergence rates for the dimension reduction of 3D ground state and
 dynamics of the GPE in the case of
 disk-shaped condensation and cigar-shaped condensation are
 reported based on our asymptotic and numerical results.
In addition, the parameter regimes in which the 3D GPE cannot be
reduced to lower dimensions are identified.

\end{abstract}



\section{Introduction}\label{s1}
\setcounter{equation}{0}

Since its realization in dilute bosonic atomic gases \cite{AEMWC,
DMADDKK, Bradley}, Bose-Einstein condensation (BEC) has been
produced and studied extensively in the laboratory \cite{CH,LS,Hall},
and has afforded an  intriguing glimpse into the macroscopic
quantum world \cite{LS}. The experimental advances in BEC have
spurred great excitement in the atomic physics community and
renewed the interest in studying the collective dynamics of
macroscopic ensembles of atoms occupying the same one-particle
quantum state. Needless to say that this dramatic progress on the
experimental front has stimulated a wave of activity on both the
theoretical and the numerical front.

 At temperatures $T$ much smaller than the critical temperature
$T_c$ \cite{LE}, a weakly interacting 
BEC is well described by the macroscopic wave
function
 $\psi = \psi({\bf x},t)$ whose evolution is governed by
a self-consistent, mean field nonlinear Schr\"{o}dinger equation
(NLSE) known as the  Gross-Pitaevskii equation (GPE)
\cite{EN,LZ,LS}
\begin{equation}
\label{gpe1} i\hbar\frac{\partial \psi({\bf x},t)}{\partial t}=
-\frac{\hbar^2}{2m}\nabla^2 \psi({\bf x},t)+ V({\bf x})\psi({\bf
x},t) +N U_0 |\psi({\bf x},t)|^2\psi({\bf x},t),
\end{equation}
where ${\bf x}=(x,y,z)$ is the spatial coordinate, $m$ is the
atomic mass, $\hbar$ is the Planck constant, $N$ is the number of
particles in the condensate, $V({\bf x})$ is an external trapping
potential. When  a harmonic trap potential is considered, $V({\bf
x})=\frac{m}{2}\left(\omega_x^2 x^2+\omega_y^2 y^2 +\omega_z^2
z^2\right)$ with $\omega_x$, $\omega_y$ and $\omega_z$ being the
trap frequencies in $x$, $y$ and $z$-direction, respectively.
Without loss of generality, we assume $\omega_x\le \omega_y\le
\omega_z$. $U_0=\frac{4\pi \hbar^2 a_s}{m}$ describes the
interaction between atoms in the condensate with $a_s$ the
$s$-wave scattering length. The
wave function is normalized according to 
 \begin{equation}
\label{norm} \int_{{\Bbb R}^3} \; |\psi({\bf x},t)|^2\;d{\bf x}=1.
\end{equation}

Following the physical literatures \cite{MMFSM,WW,WDP,LS},
introducing the rescaling: $t\to \frac{t}{\omega_x}$, ${\bf x} \to
a_0 {\bf x}$ with $a_0=\sqrt{\frac{\hbar}{m\omega_x}}$, and
$\psi\to \frac{\psi}{a_0^{3/2}}$, we get the following
dimensionless GPE under the normalization (\ref{norm}) in 3D:
\begin{equation}
\label{gpe2} i\;\frac{\partial \psi({\bf x},t)}{\partial
t}=-\frac{1}{2}\nabla^2 \psi({\bf x},t)+ V({\bf x})\psi({\bf x},t)
+ \beta\; |\psi({\bf x},t)|^2\psi({\bf x},t),
\end{equation}
where $\beta=\frac{U_0 N}{a_0^3\hbar \omega_x}=\frac{4\pi
a_sN}{a_0}$ and $V({\bf x})=\frac{1}{2}\left( x^2+\gamma_y^2 y^2+
\gamma_z^2 z^2\right)$ with  $\gamma_y =
\frac{\omega_y}{\omega_x}$ and $\gamma_z =
\frac{\omega_z}{\omega_x}$.

To find the stationary solution of (\ref{gpe2}), we write
\begin{equation}
\label{state-az} \psi({{\bf x}},t)=\phi({{\bf x}})\; e^{-i\mu t},
\end{equation}
where $\mu$ is the chemical potential of the condensate and
$\phi({{\bf x}})$ is a function independent of time. Substituting
(\ref{state-az}) into (\ref{gpe2}) gives the following equation
for $(\mu, \phi({{\bf x}}))$:
\begin{equation}
\label{nleg1} \mu\;\phi({{\bf x}}) =  -\frac{1}{2}\nabla^2
\phi({{\bf x}})+V({{\bf x}})\phi({{\bf x}}) +\beta| \phi({{\bf
x}})|^2\phi({{\bf x}}), \qquad {\bf x}\in{\Bbb R}^3,
\end{equation}
under  the normalization condition (\ref{norm}) with $\psi=\phi$.

This is a  nonlinear eigenvalue problem with a constraint and any
eigenvalue  $\mu$ can be computed from its corresponding
eigenfunction $\phi({{\bf x}})$ by
\begin{equation}
\label{cp1} \mu =  \mu(\phi) =\int_{{\Bbb
R}^3}\left[\frac{1}{2}|\nabla\phi({{\bf x}})|^2+ V({{\bf
x}})|\phi({{\bf x}})|^2+\beta|\phi({{\bf x}})|^4 \right]d{{\bf x}}
=  E(\phi)+E_{\rm int}(\phi),
\end{equation}
where
\begin{eqnarray}
&&E(\phi)=\int_{{\Bbb
R}^3}\left[\frac{1}{2}|\nabla\phi({{\bf x}})|^2+ V({{\bf
x}})|\phi({{\bf x}})|^2+\frac{\beta}{2}|\phi({{\bf x}})|^4 \right]d{{\bf x}},\\
 \label{ke1}
&&E_{\rm int}(\phi)=\frac{\beta}{2}\int_{{\Bbb R}^3} |\phi({{\bf
x}})|^4\;d{\bf x}, \qquad E_{\rm kin}(\phi) = \int_{{\Bbb R}^3}
\frac{1}{2}|\nabla\phi({{\bf x}})|^2\; d{\bf x}.
\end{eqnarray}

The ground state of a BEC is usually defined as the minimizer of
the following minimization problem:

Find $(\mu_g, \phi_g\in S)$ such that
\begin{equation}
\label{minp} E_g:=E(\phi_g) = \min_{\phi\in S} E(\phi), \qquad
\mu_g:=\mu(\phi_g)= E(\phi_g)+E_{\rm int}(\phi_g),
\end{equation}
where $S=\{\phi \ |\ \|\phi\|=1, \ E(\phi)<\infty\}$ is the  unit
sphere. The existence of a unique positive  minimizer of the
minimization  problem (\ref{minp}) is given in \cite{ERJ}.

   In an experimental setup, the trapping frequencies in
different directions can be very different. Especially, disk-shaped
 and cigar-shaped condensation were observed in
experiments \cite{BP,FSLS,LSe,LS}. 
The 3D GPE (\ref{gpe2}) is formally reduced to 2D GPE in
disk-shaped condensation and to 1D GPE in cigar-shaped
condensation in the literatures
\cite{FSLS,Lieb, Sei,LSe,LS,ERJ,PN,WW,WDP,Bao,Sala}. Mathematical and 
numerical
justification for the dimension reduction of 3D GPE is only
available in the weakly interaction regime, i.e. $\beta=o(1)$
\cite{BaoM, Ben}. Unfortunately, in the intermediate or strong
interaction regime, no results are available. The aim of this
paper is to study
 dimension reduction of the 3D GPE
(\ref{gpe2}) under different limiting parameter regimes. In
addition, we will provide convergence rate for the dimension
reduction based on our asymptotic and numerical results.

   The paper is organized as follows.  In section \ref{s2} we
study dimension reduction for the ground state of the 3D GPE
(\ref{nleg1}) with replusive interaction $\beta>0$
and provide convergence rates. 
Convergence rate for dimension reduction of 3D time-dependent GPE
(\ref{gpe2}) is reported in section \ref{s3}.
 Finally  some conclusions are
drawn in section \ref{sc}.

\section{Dimension reduction for ground  states}\label{s2}
\setcounter{equation}{0}

  In this section, we will discuss dimension reduction for
BEC  ground states and derive approximate ground states as well as
their energy and chemical potential for the  3D GPE (\ref{nleg1})
under different parameter regimes of $\beta$, $\gamma_y$ and
$\gamma_z$.

\subsection{Isotropic shaped condensation}

In the case of isotropic condensates, i.e.
$\gamma_y=O(1)$ and $\gamma_z=O(1)$ ($\Longleftrightarrow$
$\omega_y\approx \omega_x$ and $\omega_z\approx \omega_x$), there
are three typical regimes:

\bigskip

\noindent Regime I. Weakly interacting regime, i.e. $\beta=o(1)$,
the ground state is well approximated by the harmonic oscillator
ground state \cite{MSM,FSLS,LS,WW}:
\begin{eqnarray}
\label{gsa1}
&&\phi_g({\bf x}) \approx \phi_{\rm ho}(x,y,z)=
\frac{(\gamma_y\gamma_z)^{1/4}}{\pi^{3/4}}
e^{-\frac{x^2 +\gamma_{_y} y^2 +\gamma_{_z} z^2}{2}}, \qquad {\bf x}\in
{\Bbb R}^3, \\
\label{ega1} &&E_g\approx E(\phi_{\rm
ho})=\frac{1}{2}\left(1+\gamma_y+\gamma_z\right)+O(\beta),
\qquad |\beta|\ll 1,\\
\label{muga1} &&\mu_g \approx \mu(\phi_{\rm ho})=
\frac{1}{2}\left(1+\gamma_y+\gamma_z\right)+O(\beta).
\end{eqnarray}

\bigskip

\noindent Regime II. Intermediate  interacting regime,
$\beta=O(1)$, the ground state can be obtained by solving the 3D
minimization problem (\ref{minp}). Different numerical methods
were proposed in the literatures for computing the ground states
\cite{WW,WD,Bao,MMFSM,MSM}.

\bigskip

\noindent Regime III. Strong interacting regime,
$\beta\gg1$, noticing (\ref{Egtfra}) and (\ref{TFcp}),
 the ground state is approximated by the Thomas-Fermi (TF)
 approximation \cite{MSM,FSLS,LS,WW}:
\begin{eqnarray}
\label{gsa2} &&\phi_g({\bf x}) \approx \phi_g^{\rm
TF}(\bx)=\left\{\begin{array}{ll} \sqrt{\left(\mu_g^{\rm
TF}-V({\bf x})\right)/\beta},
 &\quad V({\bf x}) < \mu_g^{\rm TF}, \\
0, &\quad \hbox{otherwise},\\
\end{array}\right. \qquad  \\
&&\mu_g^{\rm TF}=\frac{1}{2}\left(\frac{15 \beta \gamma_y
\gamma_z}{4\pi}\right)^{2/5}, \qquad  \beta\gg1,\\
 \label{ega2}
&&E_g \approx \frac{5}{7}\mu_g^{\rm TF}+
\frac{\tilde{C}_3}{\beta^{2/5}}\left(\ln \beta +G_3\right)
=\frac{5}{7}\mu_g^{\rm
TF}+O\left(\frac{\ln\beta}{\beta^{2/5}}\right), \\
\label{muga2}
&&\mu_g \approx \mu_g^{\rm TF}+\frac{\tilde{C}_3}{\beta^{2/5}}
\left(\ln \beta +G_3\right)
=\mu_g^{\rm TF}+O\left(\frac{\ln\beta}{\beta^{2/5}}\right).
\end{eqnarray}

 For $\gamma_y = \gamma_z=1$, (\ref{ega2}) and (\ref{muga2})
were confirmed numerically in \cite{WW}.

\subsection{Disk-shaped condensation}

In the case of disk shaped condensates, i.e. $\gamma_y=O(1)$ and
$\gamma_z\gg 1$ ($\Longleftrightarrow$  $\omega_y\approx \omega_x$
and $\omega_z\gg \omega_x$), we set
\begin{equation}
\label{trans32d}
\mu_g \approx \mu +\frac{\gamma_z}{2}, \qquad
 \phi_g({\bf x}) \approx \phi(x,y) \phi_{\rm ho}(z) \quad \hbox{with} \
\phi_{\rm ho}(z)= \frac{\gamma_z^{1/4}}{\pi^{1/4}}
e^{-\frac{\gamma_{_z} z^2}{2}}.
\end{equation}
Plugging (\ref{trans32d}) into (\ref{nleg1}), multiplying both sides by
$\phi_{\rm ho}(z)$ and integrating over $z\in(-\infty,\infty)$, we
get
\begin{equation}
\label{nleg6}
\mu\; \phi(x,y) = -\frac{1}{2} \Delta \phi(x,y) +V_2(x,y)
\phi(x,y) + \beta_2^a\;|\phi(x,y)|^2\phi(x,y), \qquad (x,y)\in {\Bbb R}^2,
\end{equation}
where \[V_2(x,y)= \frac{1}{2}\left(x^2 +\gamma_y^2
y^2\right),\qquad  \beta_2^a = \beta\int_{-\infty}^\infty |\phi_{\rm
ho}(z)|^4\;dz=\beta \sqrt{\frac{\gamma_z}{2\pi}}.\]
 Using the
results in Appendix A for 2D GPE, again we get approximate ground
state in three typical regimes:

\bigskip

\noindent Regime I. Weakly interacting regime, i.e.
$\beta_2^a=\beta\sqrt{\gamma_z/2\pi}=o(1)$, the ground state is
approximated by the harmonic oscillator ground state:
\begin{eqnarray}
\label{gsa3}
&&\phi_g({\bf x}) \approx \phi_{\rm ho}(x,y) \phi_{\rm ho}(z)=
\phi_{\rm ho}(x,y,z), \qquad {\bf x}\in
{\Bbb R}^3, \\
\label{ega3}
&&E_g\approx \frac{\gamma_z}{2}+\frac{1+\gamma_y}{2}+
 O(\beta_2^a)=\frac{\gamma_z}{2}+\frac{1+\gamma_y}{2}+O\left(
\beta \gamma_z^{1/2}\right), \\
\label{muga3} &&\mu_g \approx
\frac{\gamma_z}{2}+\frac{1+\gamma_y}{2}+ O(\beta_2^a)=
\frac{\gamma_z}{2}+\frac{1+\gamma_y}{2}+O\left( \beta
\gamma_z^{1/2}\right), \ \gamma_z\gg1\; \& \;\beta_2^a=o(1).
\qquad\quad
\end{eqnarray}

\bigskip

\noindent Regime II. Intermediate  interacting
regime, i.e. $\beta_2^a=O(1)$, 
the ground state can
be approximated by
\begin{equation}
\label{gsa4}
\phi_g({\bf x}) \approx \phi_g^{\rm DS}({\bf x}):=\phi_g^{\rm 2D}(x,y)
\phi_{\rm ho}(z), \qquad {\bf x}\in {\Bbb R}^3.
\end{equation}

\begin{eqnarray}
\label{ega4}
&&E_g\approx E_g^{\rm DS}:=E(\phi_g^{\rm 2D}(x,y)\phi_{\rm ho}(z))=
\frac{\gamma_z}{2}+ E_{\rm 2D}(\phi_g^{\rm 2D}):= \frac{\gamma_z}{2}+
 E_g^{\rm 2D}, \\
\label{muga4}
&&\mu_g\approx\mu_g^{\rm DS}:=\mu(\phi_g^{\rm 2D}(x,y)\phi_{\rm ho}(z))=
\frac{\gamma_z}{2}+\mu_{\rm 2D}(\phi_g^{\rm 2D}):=
\frac{\gamma_z}{2}+ \mu_g^{\rm 2D},
\end{eqnarray}
where
\begin{eqnarray*}
&&E_g^{\rm 2D}=\int_{{\Bbb R}^2} \left[\frac{1}{2}
|\nabla \phi_g^{\rm 2D}|^2 + V_2(x,y) | \phi_g^{\rm 2D}|^2
+\frac{\beta_2^a}{2}| \phi_g^{\rm 2D}|^4\right]\;dxdy, \\
&&\mu_g^{\rm 2D}=\int_{{\Bbb R}^2} \left[\frac{1}{2}
|\nabla \phi_g^{\rm 2D}|^2 + V_2(x,y) | \phi_g^{\rm 2D}|^2
+\beta_2^a| \phi_g^{\rm 2D}|^4\right]\;dxdy.
\end{eqnarray*}
Here $\phi_g^{\rm 2D}$, $E_g ^{\rm 2D}$ and $\mu_g^{\rm 2D}$
are the ground state, energy and chemical potential of the 2D
problem (\ref{nleg6}). In this case, one only needs to  solve a 2D problem
numerically and thus significantly 
save computational time, memory and cost.

 To verify (\ref{gsa4}), (\ref{ega4}) and (\ref{muga4}) numerically,
   Table 1 list the error
$\|\phi_g({\bf x})-\phi_g^{\rm DS}({\bf x}) \|_{L^2}$, and  Figure
1 shows the error and $|E_g - E_g^{\rm DS}|$, with $\gamma_y=1$
for different $\beta$ and $\gamma_z$. Here and in the following,
the ground state $\phi_g$ is computed numerically by the
continuous normalized gradient flow (CNGF) with a backward Euler
finite difference (BEFD) discretization \cite{WD}.

\begin{table}[t!]
\begin{center}
\begin{tabular}{lcccc}\hline
$1/\gamma_z$ & $1/25$ &$1/100$ &$1/400$   &$1/1600$ \\
\hline

$\beta=1$  &2.23E-3 &9.97E-4 &4.17E-4 &1.68E-4 \\
rate &  &0.58 &0.63 &0.65 \\ \hline

 $\beta=10$  &1.23E-2 &4.72E-3 &1.74E-3 &5.80E-4 \\
rate &  &0.70 &0.72 &0.80 \\ \hline

 $\beta=100$  &4.46E-2 &1.64E-2 &5.93E-3 &2.13E-3 \\
rate &  &0.72 &0.73 &0.74 \\ \hline

$\beta=1000$  &1.38E-1 &5.31E-2 &1.94E-2 &6.99E-3 \\
rate &  &0.69 &0.73 &0.74 \\ \hline

$\beta=10000$  &3.46E-1 &1.58E-1 &6.19E-2 &2.27E-2 \\
rate &  &0.56 &0.68 &0.72 \\ \hline

\end{tabular}
\end{center}
\caption{Error analysis of
$\|\phi_g(\cdot)-\phi_g^{\rm DS}(\cdot)\|_{L^2}$
for the ground state in 3D with a disk-shaped trap.}
\end{table}






 \begin{figure}[t!]
\centerline{(a)\psfig{figure=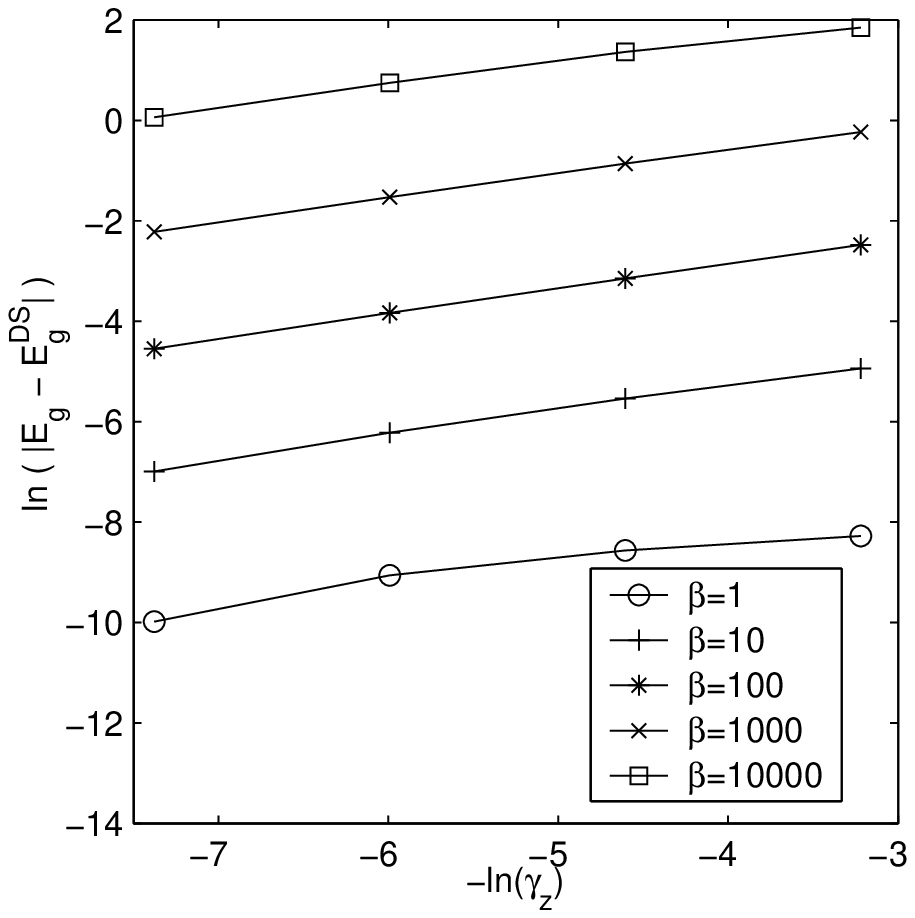,height=7cm,width=6.5cm,angle=0}
\quad (b)\psfig{figure=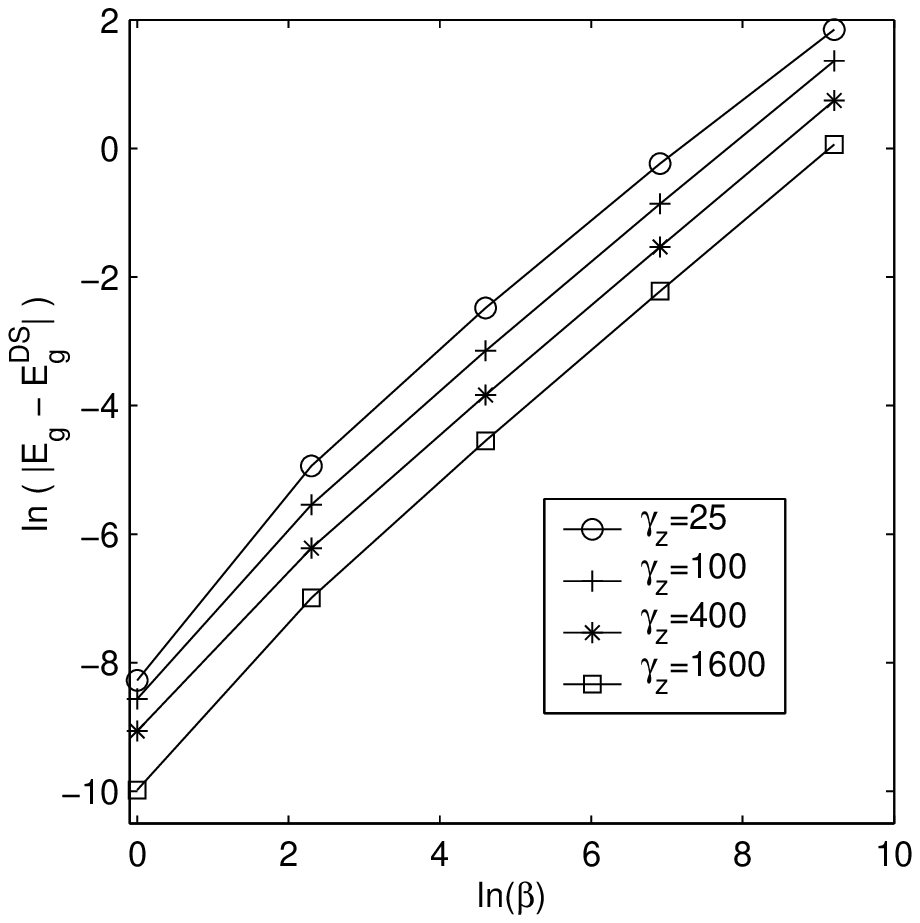,height=7cm,width=6.5cm,angle=0}  }

Figure 1: Convergence rate of $|E_g - E_g^{\rm DS}|$ in 3D with a
disk-shaped trap: (a) With respect to $\gamma_z$; (b) with respect
to $\beta$.
\end{figure}

 From Tab. 1, Fig. 1 and additional numerical
results \cite{GeY}, when $\gamma_z\gg1$,
$\beta_2^a=\beta\sqrt{\gamma_z/2\pi}=O(1)$ or $\gg1$, and
$\beta\gamma_z^{-3/2}=o(1)$, we can draw the following conclusion:
\begin{eqnarray*}
&&\|\phi_g-\phi_g^{\rm DS}\|_{L^2}=
O\left(\frac{\beta^{1/2}\; \ln\gamma_z}{\gamma_z^{3/4}}\right), \quad
\|\phi_g^2-(\phi_g^{\rm DS})^2\|_{L^1}
=O\left(\frac{\beta^{1/2}\; \ln\gamma_z}{\gamma_z^{3/4}}\right), \\
&&|E_g-E_g^{\rm DS}|=
O\left(\frac{\beta\; \ln\gamma_z}{\gamma_z^{1/2}}\right),
\quad
 |\mu_g-\mu_g^{\rm DS}|=
O\left(\frac{\beta\; \ln\gamma_z}{\gamma_z^{1/2}}\right).
\end{eqnarray*}
On the contrary, when $\gamma_z\gg1$,
$\beta_2^a=\beta\sqrt{\gamma_z/2\pi}=O(1)$ or $\gg1$, and
$\beta\gamma_z^{-3/2}=O(1)$ or $\gg1$, the errors do not decrease
when $\gamma_z$ increases. This suggests that one cannot reduce the
3D GPE (\ref{nleg1}) to 2D GPE (\ref{nleg6})  when $\beta\gg1$ and
$\gamma_z\gg1$ with  $\beta\gamma_z^{-3/2}=O(1)$ or $\gg1$.

\bigskip

\noindent Regime III. Strong interacting regime,
$\beta_2^a=\beta\sqrt{\gamma_z/2\pi}\gg1$, noticing
(\ref{faenergy}),
 the ground state is approximated by the multiplication of
the TF approximation in $xy$-plane and the harmonic oscillator
approximation in $z$-direction:
\begin{equation}
\label{gsa5}
\phi_g({\bf x}) \approx \phi_g^{\rm TF1}({\bf x}):=
\phi_{\rm 2D}^{\rm TF}(x,y)\phi_{\rm ho}(z),  \qquad {\bf x}\in {\Bbb R}^3,
\end{equation}
where
\begin{eqnarray}
\label{phidsTF}
 &&\phi_{\rm 2D}^{\rm TF}(x,y)=
\left\{\begin{array}{ll}
\sqrt{\left(\mu_{\rm 2D}^{\rm TF}-V_2(x,y)\right)/\beta_2^a},
 &V_2(x,y)  < \mu_{\rm 2D}^{\rm TF}, \\
0 &\hbox{otherwise},\\
\end{array}\right.  \\
\label{mu2D7}
&&\mu_{\rm 2D}^{\rm TF} = \left(\frac{\beta_2^a \gamma_y}{\pi}\right)^{1/2}
=\frac{\beta^{1/2}\gamma_y^{1/2} \gamma_z^{1/4}}{2^{1/4}\pi^{3/4}}.
\end{eqnarray}
Plugging (\ref{gsa4}), (\ref{nleg6}), (\ref{Egtfra}) with $d=2$,
(\ref{mu2D7}), (\ref{kenergy}) with $d=2$ and
$\beta_2=\beta_2^a=\beta\sqrt{\frac{\gamma_z}{2\pi}}$
 into (\ref{ega4}), we get the approximate energy
\begin{eqnarray}
\label{ega5}
E_g&=&E(\phi_g)=E(\phi_g^{\rm 2D}(x,y)\phi_{\rm ho}(z))+O\left(\frac{\beta
\ln \gamma_z}{\gamma_z^{1/2}} \right)\nonumber\\
&=&\frac{\gamma_z}{2}+E_{\rm 2D}(\phi_g^{\rm 2D})+O\left(\frac{\beta
\ln \gamma_z}{\gamma_z^{1/2}}\right)=
\frac{\gamma_z}{2} + E_g^{\rm 2D}+O\left(\frac{\beta
\ln \gamma_z}{\gamma_z^{1/2}}\right)   \nonumber \\
&\approx&\frac{\gamma_z}{2} + \frac{2}{3} \left(\frac{\beta_2^a \gamma_y}
{\pi}\right)^{1/2} + \frac{\tilde{C}_2}{(\beta_2^a)^{1/2}}
\left(\ln \beta_2^a +G_2\right)+O\left(\frac{\beta
\ln \gamma_z}{\gamma_z^{1/2}}\right) \nonumber \\
&\approx&\frac{\gamma_z}{2} +\frac{2^{3/4}\gamma_y^{1/2}(\beta^2
\gamma_z)^{1/4}}{3\pi^{3/4}} +
\frac{\tilde{C}_2 (2\pi)^{1/4}}{2(\beta^2\gamma_z)^{1/4}}\left[
\ln (\beta^2 \gamma_z)+2G_2 -\ln 2\pi\right] +O\left(\frac{\beta
\ln \gamma_z}{\gamma_z^{1/2}}\right)\nonumber \\
&=&E_g^{\rm TF1} + O\left(\frac{\ln (\beta^2 \gamma_z)}{
(\beta^2 \gamma_z)^{1/4}}+\frac{\beta
\ln \gamma_z}{\gamma_z^{1/2}}\right),
\end{eqnarray}
where
\begin{equation}
E_g^{\rm TF1}=\frac{\gamma_z}{2} +\frac{2^{3/4}\gamma_y^{1/2}(\beta^2
\gamma_z)^{1/4}}{3\pi^{3/4}}.
\end{equation}
Similarly, we get the approximate chemical potential:
\begin{equation}
\label{muga5}
\mu_g\approx\mu_g^{\rm TF1} + O\left(\frac{\ln (\beta^2 \gamma_z)}{
(\beta^2 \gamma_z)^{1/4}}+\frac{\beta
\ln \gamma_z}{\gamma_z^{1/2}} \right),
\end{equation}
where
\begin{equation}
\mu_g^{\rm TF1}=\frac{\gamma_z}{2} +\frac{\gamma_y^{1/2}(\beta^2
\gamma_z)^{1/4}}{2^{1/4}\pi^{3/4}}.
\end{equation}

 To verify (\ref{gsa5}), (\ref{ega5}) and (\ref{muga5}) numerically,
   Tables 2 and 3 list the errors
$\|\phi_g({\bf x})-\phi^{\rm TF1}({\bf x})\|_{L^2}$ and
$\|\phi_g({\bf x})-\phi^{\rm TF1}({\bf x})\|_{L^\infty}$, respectively,
  and  Figure 2
shows the error $|E_g - E_g^{\rm TF1}|$, with $\gamma_y=1$ for
different $\beta$ and $\gamma_z$.

\begin{table}[t!]
\begin{center}
\begin{tabular}{lcccc}\hline
$1/\gamma_z$ & $1/25$ &$1/100$ &$1/400$   &$1/1600$ \\
\hline

$\beta=1$  &5.45E-1 &4.44E-1 &3.57E-1 &2.88E-1 \\
rate &  &0.15 &0.16 &0.15 \\ \hline

 $\beta=10$  &2.66E-1 &2.15E-1 &1.74E-1 &1.40E-1 \\
rate &  &0.15 &0.16 &0.15 \\ \hline

 $\beta=100$  &1.29E-1 &1.03E-1 &8.43E-2 &6.77E-2 \\
rate &  &0.16 &0.14 &0.16 \\ \hline

 $\beta=1000$  &1.40E-1 &6.49E-2 &4.12E-2 &3.19E-2 \\
rate &  &0.55 &0.33 &0.18 \\ \hline


\end{tabular}
\end{center}
\caption{Error analysis of
$\|\phi_g-\phi^{\rm TF1}\|_{L^2}$
for the ground state in 3D with a disk-shaped trap.}
\end{table}

\begin{table}[t!]
\begin{center}
\begin{tabular}{lcccc}\hline
$1/\gamma_z$ & $1/25$ &$1/100$ &$1/400$   &$1/1600$ \\
\hline

$\beta=1$  &0.3880 &0.3148 &0.3107 &0.3749 \\

 $\beta=10$  &3.845E-2 &4.589E-2 &5.337E-2 &6.198E-2 \\

 $\beta=100$  &6.368E-3 &7.048E-3 &8.597E-3 &9.757E-3 \\
\hline
\end{tabular}
\end{center}
\caption{Error analysis of
$\|\phi_g-\phi^{\rm TF1}\|_{L^\infty}$
for the ground state in 3D with a disk-shaped trap.}
\end{table}






 \begin{figure}[t!]
\centerline{(a)\psfig{figure=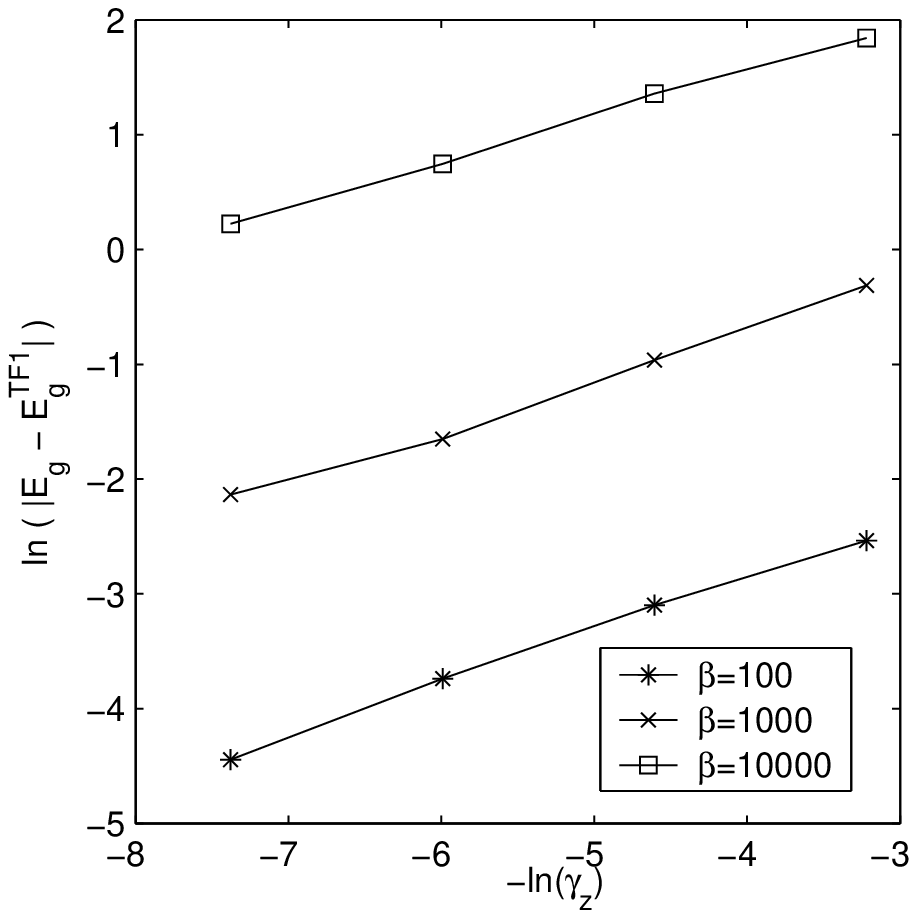,height=7cm,width=6.5cm,angle=0}
\quad (b)\psfig{figure=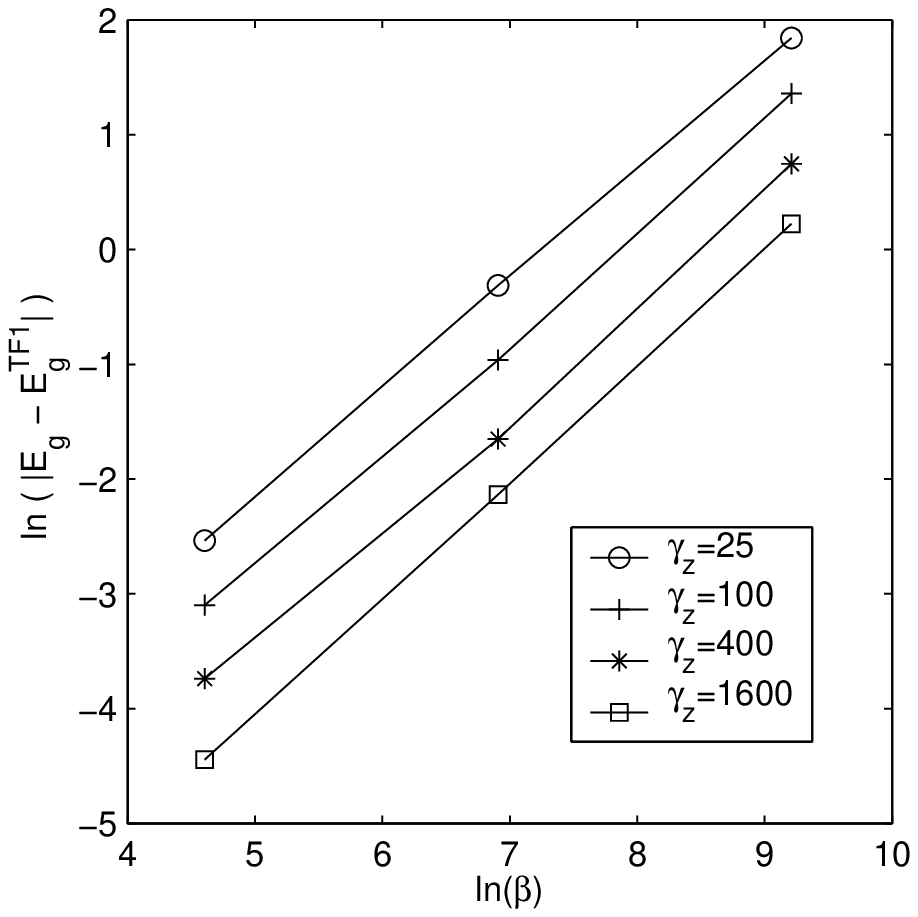,height=7cm,width=6.5cm,angle=0}  }

Figure 2: Convergence rate of $|E_g - E_g^{\rm TF1}|$ in 3D with a
disk-shaped trap: (a) With respect to $\gamma_z$; (b) with respect
to $\beta$.
\end{figure}

 From Tabs. 2\&3, Fig. 2 and  additional numerical results in 
\cite{GeY}, when $\gamma_y=O(1)$, $\gamma_z\gg1$,
$\beta_2^a=\beta\sqrt{\gamma_z/2\pi}\gg1$ and
 $\beta \gamma_z^{-3/2}=o(1)$, we can draw the
following conclusion:
\begin{eqnarray*}
&&\|\phi_g-\phi^{\rm TF1}\|_{L^2}=
O\left(\frac{C(\beta)\;\ln\gamma_z}{\gamma_z^{1/4}}\right), \quad
\|\phi_g^2-(\phi^{\rm TF1})^2\|_{L^1}
=O\left(\frac{C(\beta)\; \ln\gamma_z}{\gamma_z^{1/4}}\right), \\
&&|E_g-E_g^{\rm TF1}|=
O\left(\frac{C(\beta)\;\ln\gamma_z}{\gamma_z^{1/4}}\right), \qquad
|\mu_g-\mu_g^{\rm TF1}|= O\left(\frac{C(\beta)\;
\ln\gamma_z}{\gamma_z^{1/4}}\right),
\end{eqnarray*}
where $C(\beta)$ depends on $\beta$. These results confirm the
asymptotic results (\ref{ega5}) and (\ref{muga5}). Furthermore,
the  additional numerical results in \cite{WW,GeY} indicate that
$\phi^{\rm TF1}({\bf x})$ does not converges pointwise to the
ground state
 $\phi_g({\bf x})$ when $\gamma_z\to \infty$ and $\beta>0$ 
(cf. Tab. 3).

\subsection{Cigar-shaped condensation}

In the case of cigar shaped condensates, i.e. $\gamma_y\gg1$ and
$\gamma_z\gg 1$ ($\Longleftrightarrow$ $\omega_y\gg\omega_x$ and
$\omega_z\gg \omega_x$), we set
\begin{equation}
\label{trans31d}
\mu_g \approx \mu + \frac{\gamma_y+\gamma_z}{2},
\quad \phi_g({\bf x}) \approx \phi(x) \phi_{\rm ho}(y,z), \quad
\phi_{\rm ho}(y,z)=\frac{(\gamma_y\gamma_z)^{1/4}}{\pi^{1/2}}
e^{-\frac{\gamma_{_y} y^2 +\gamma_{_z} z^2}{2}}.
\end{equation}
Plugging (\ref{trans31d}) into (\ref{nleg1}), multiplying both sides by
$\phi_{\rm ho}(y,z)$ and integrating over $(y,z)\in{\Bbb R}^2$, we
get
\begin{equation}
\label{nleg8}
\mu\; \phi(x) = -\frac{1}{2} \phi_{xx} +V_1(x)\phi
+ \beta_1^a |\phi|^2\phi, \qquad -\infty<x<\infty,
\end{equation}
where
\[V_1(x)= \frac{x^2}{ 2}, \qquad \beta_1^a =
 \beta\int_{{\Bbb R}^2} |\phi_{\rm ho}(y,z)|^4\;dydz=\frac{\beta
\sqrt{\gamma_y \gamma_z}}{2\pi}.\]
 Using the results in the
Appendix A for 1D GPE, again we get approximate ground state in
three typical regimes:

\bigskip

\noindent Regime I. Weakly interacting regime, i.e.
$\beta_1^a=\beta\sqrt{\gamma_y\gamma_z}/2\pi=o(1)$, the ground
state is approximated by the harmonic oscillator ground state:
\begin{eqnarray}
\label{gsa7}
&&\phi_g({\bf x}) \approx \phi_{\rm ho}(x)\phi_{\rm ho}(y,z)=
\phi_{\rm ho}(x,y,z), \quad {\bf x}\in
{\Bbb R}^3, \quad
 \gamma_y\gg1\;\&\;\gamma_z\gg1\;\&\;\beta_1^a=o(1), \qquad \quad \\
\label{ega7}
&&E_g\approx \frac{\gamma_y+\gamma_z}{2}+\frac{1}{2}+
 O(\beta_1^a)=\frac{\gamma_y+\gamma_z}{2}+\frac{1}{2}+
 O\left(\beta (\gamma_y \gamma_z)^{1/2}\right), \\
\label{muga7}
&&\mu_g \approx \frac{\gamma_y+\gamma_z}{2}+\frac{1}{2}+
O(\beta_1^a)=\frac{\gamma_y+\gamma_z}{2}+\frac{1}{2}
+O\left(\beta (\gamma_y \gamma_z)^{1/2}\right).
\end{eqnarray}

\bigskip

\noindent Regime II. Intermediate interacting
regime, i.e. $\beta_1^a=\beta\sqrt{\gamma_y\gamma_z}/2\pi=O(1)$, 
the ground state can be approximated by
\begin{equation}
\label{gsa8}
\phi_g({\bf x}) \approx \phi_g^{\rm CS}({\bf x}):=\phi_g^{\rm 1D}(x)
\phi_{\rm ho}(y,z), \qquad {\bf x}\in
{\Bbb R}^3,\\
\end{equation}
\begin{eqnarray}
\label{ega8}
E_g&\approx&E_g^{\rm CS}:=E(\phi_g^{\rm 1D}(x)\phi_{\rm ho}(y,z))=
\frac{\gamma_y+\gamma_z}{2}+ E_{\rm 1D}(\phi_g^{\rm 1D}):=
\frac{\gamma_y+\gamma_z}{2}+E_g^{\rm 1D},\qquad \quad \\
\label{muga8}
\mu_g&\approx&\mu_g^{\rm CS}:=\mu(\phi_g^{\rm 1D}(x) \phi_{\rm ho}(y,z))=
\frac{\gamma_y+\gamma_z}{2}  + \mu_{\rm 1D}(\phi_g^{\rm 1D}):=
\frac{\gamma_y+\gamma_z}{2}  + \mu_g^{\rm 1D}, \qquad
\end{eqnarray}
where
\begin{eqnarray*}
&&E_g^{\rm 1D}=
\int_{-\infty}^\infty \left[
\frac{1}{2}\left|\frac{d\phi_g^{\rm 1D}(x)}{dx}\right|^2 +V_1(x)
|\phi_g^{\rm 1D}(x)|^2 + \frac{\beta_1^a}{2}
|\phi_g^{\rm 1D}(x)|^4\right]dx, \\
&&\mu_g^{\rm 1D}=
\int_{-\infty}^\infty \left[
\frac{1}{2}\left|\frac{d\phi_g^{\rm 1D}(x)}{dx}\right|^2 +V_1(x)
|\phi_g^{\rm 1D}(x)|^2 + \beta_1^a
|\phi_g^{\rm 1D}(x)|^4\right]dx.
\end{eqnarray*}
Here $\phi_g^{\rm 1D}$, $E_g ^{\rm 1D}$ and $\mu_g^{\rm 1D}$
are the ground state, energy and chemical potential  of the 1D
problem (\ref{nleg8}). In this case, one only needs to solve a 1D problem
numerically and thus significantly save 
computational time, memory and cost.

 To verify (\ref{gsa4}), (\ref{ega4}) and (\ref{muga4}) numericallly,
   Table 4 lists the error
$\|\phi_g({\bf x})-\phi_g^{\rm CS}({\bf x})\|_{L^2}$, and  Figure
3 shows the error $\frac{|E_g - E_g^{\rm CS}|}{E_g}$, for
different $\beta$ and $\gamma:=\gamma_y=\gamma_z$.

\begin{table}[t!]
\begin{center}
\begin{tabular}{lccccc}\hline
$1/\gamma$ & $1/12.5$ &$1/25$ &$1/50$   &$1/100$  &$1/200$\\
\hline

$\beta=25$  &0.1512 &0.1283 &0.1076 &0.0895 &0.074\\
rate &  &0.24 &0.25 &0.27 &0.28 \\ \hline

$\beta=50$  &0.2232 &0.1914 &0.1162 &0.1363 &0.1136\\
rate &  &0.22 &0.24 &0.25 &0.26 \\ \hline

$\beta=100$  &0.3150 &0.2742 &0.2357 &0.2006 &0.1692\\
rate &  &0.20 &0.22 &0.23 &0.25 \\ \hline

$\beta=200$  &0.4228 &0.3740 &0.3269 &0.2826 &0.2418\\
rate &  &0.18 &0.19 &0.21 &0.22 \\ \hline

$\beta=400$  &0.5389 &0.4851 &0.4316 &0.3798 &0.3309\\
rate &  &0.15 &0.17 &0.18 &0.20 \\ \hline

\end{tabular}
\end{center}
\caption{Error analysis of
$\|\phi_g-\phi_g^{\rm CS}\|_{L^2}$
for the ground state in 3D with a cigar-shaped trap.}
\end{table}

 \begin{figure}[t!]
\centerline{(a)\psfig{figure=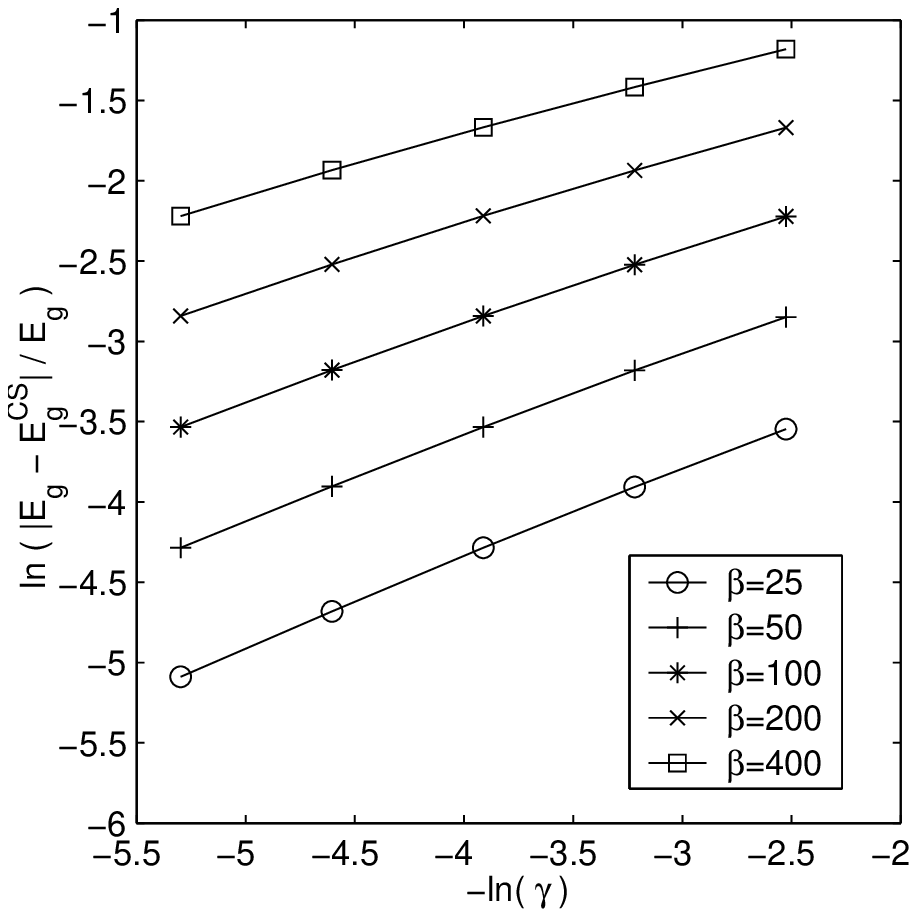,height=7cm,width=6.5cm,angle=0}
\quad (b)\psfig{figure=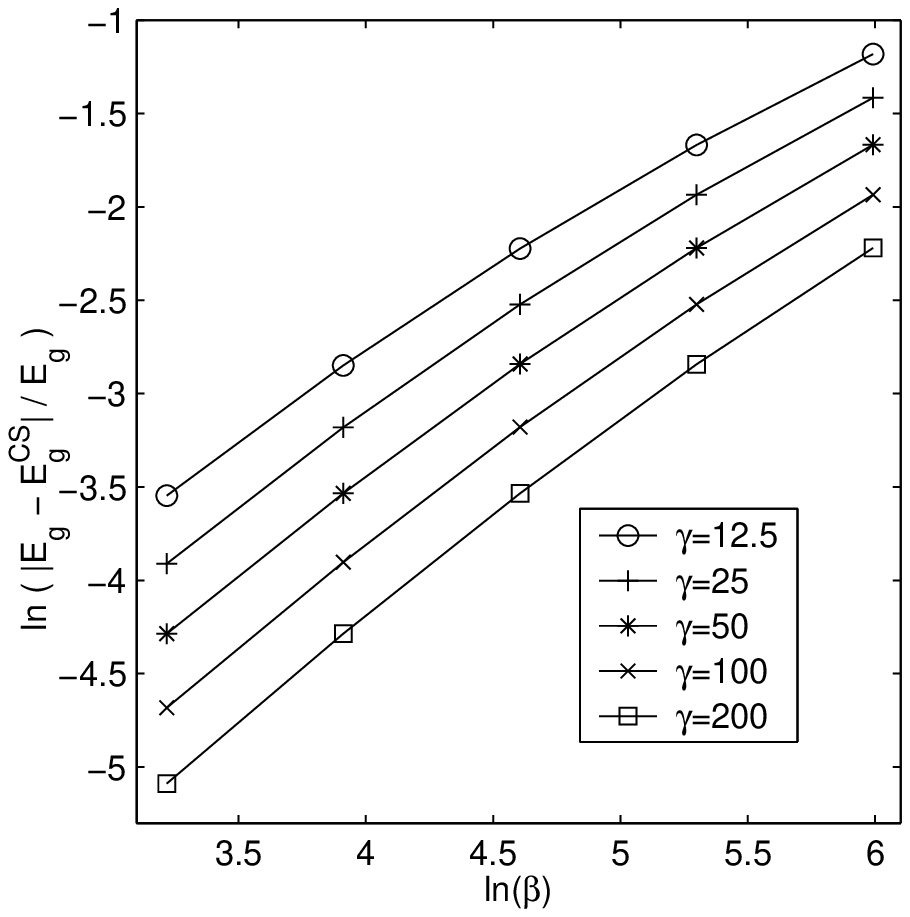,height=7cm,width=6.5cm,angle=0}  }

Figure 3: Convergence rate of $\frac{|E_g - E_g^{\rm CS}|} {E_g}$
in 3D with a cigar-shaped trap: (a) With respect to $\gamma$; (b)
with respect to $\beta$.

\end{figure}

 From Tab. 4, Fig. 3 and  additional numerical results
in \cite{GeY}, when $\gamma=\gamma_y=\gamma_z\gg1$,
$\beta_1^a=\beta\gamma/2\pi=O(1)$ or $\gg1$, and
$\beta\gamma^{-1}=o(1)$,
 we can draw the  following conclusion:
\begin{eqnarray*}
&&\|\phi_g^2-(\phi_g^{\rm CS})^2\|_{L^1}=
O\left(\frac{\beta^{1/3}\; \ln\gamma}{\gamma^{1/3}}\right), \qquad
\|\phi_g-\phi_g^{\rm CS}(x)\|_{L^2}=
O\left(\frac{\beta^{1/3}\; \ln\gamma}{\gamma^{1/3}}\right), \\
&&|E_g-E_g^{\rm CS}|=
O\left(\beta\; \gamma^{1/3} \ln\gamma\right), \qquad
\frac{|E_g-E_g^{\rm CS}|}{E_g}=
O\left(\frac{\beta^{1/3}\; \ln\gamma}{\gamma^{2/3}}\right), \\
&&|\mu_g-\mu_g^{\rm CS}|=
O\left(\beta\; \gamma^{1/3} \ln\gamma\right), \qquad
\frac{|\mu_g-\mu_g^{\rm CS}|}{\mu_g}=
O\left(\frac{\beta^{1/3}\; \ln\gamma}{\gamma^{2/3}}\right).
\end{eqnarray*}
On the contrary, when $\gamma=\gamma_y=\gamma_z\gg1$,
$\beta_1^a=\beta\gamma/2\pi=O(1)$ or $\gg1$, and
$\beta\gamma^{-1}=O(1)$ or $\gg1$, the errors do not decrease when
$\gamma$ increases. This suggests that one cannot reduce the 3D GPE
(\ref{nleg1}) to 1D GPE (\ref{nleg8}) when $\beta\gg1$ and
$\gamma\gg1$ with  $\beta\gamma^{-1}=O(1)$ or $\gg1$.

\bigskip

\noindent Regime III. Strong  interacting regime,
$\beta_1^a=\beta\sqrt{\gamma_y\gamma_z}/2\pi\gg1$, noticing
(\ref{faenergy}) with $d=1$,
 the ground state in 3D is approximated by the multiplication of
the TF approximation in $x$-direction and the harmonic oscillator
approximation in $yz$-plane:
\begin{equation}
\label{gsa9}
\phi_g({\bf x}) \approx \phi_g^{\rm TF2}({\bf x}):=
\phi_{\rm 1D}^{\rm TF}(x) \phi_{\rm ho}(y,z),  \qquad {\bf x}\in {\Bbb R}^3,
\end{equation}
where
\begin{eqnarray}
\label{phicsTF}
&&\phi_{\rm 1D}^{\rm TF}(x) =\left\{\begin{array}{ll}
\sqrt{\left(\mu_{\rm 1D}^{\rm TF}-x^2/2\right)/\beta_1^a},
 &x^2 < 2\mu_{\rm 1D}^{\rm TF}, \\
0 &\hbox{otherwise},\\
\end{array}\right. \\
\label{mug91D}
&&\mu_{\rm 1D}^{\rm TF} = \frac{1}{2}\left(\frac{3\beta_1^a}{2}\right)^{2/3}
=\frac{(3\beta)^{2/3}(\gamma_y\gamma_z)^{1/3}}{2(4\pi)^{2/3}}.
\end{eqnarray}
Plugging (\ref{gsa8}), (\ref{nleg8}), (\ref{Egtfra}) with $d=1$,
(\ref{mug91D}), (\ref{kenergy}) with $d=1$ and $\beta_1=\beta_1^a$
 into (\ref{ega4}), we get the approximate energy:
\begin{eqnarray}
\label{ega9}
E_g&=&E(\phi_g)=E_g(\phi_g^{\rm 1D}(x) \phi_{\rm ho}(y,z))
+O\left(\beta \gamma_y^{1/3} \ln \gamma_y\right) \nonumber\\
&=&\frac{\gamma_y+\gamma_z}{2}+E_{\rm 1D}(\phi_g^{\rm 1D})
+O\left(\beta \gamma_y^{1/3} \ln \gamma_y\right)
=\frac{\gamma_y+\gamma_z}{2} + E_g^{\rm 1D}
+O\left(\beta \gamma_y^{1/3} \ln \gamma_y\right) \nonumber \\
&\approx&\frac{\gamma_y+\gamma_z}{2}
+ \frac{3}{5} \frac{1}{2} \left(\frac{3\beta_1^a}
{2}\right)^{2/3} + \frac{\tilde{C}_1}{(\beta_1^a)^{2/3}}
\left(\ln \beta_1^a +G_1\right)+O\left(\beta \gamma_y^{1/3}
\ln \gamma_y\right)
\nonumber \\
&\approx&E_g^{\rm TF2} + O\left(\beta \gamma_y^{1/3} \ln \gamma_y\right),
\end{eqnarray}
where
\begin{equation}
E_g^{\rm TF2}=\frac{\gamma_y+\gamma_z}{2} +\frac{3^{5/3} (\beta^2
\gamma_y\gamma_z)^{1/3}}{10(4\pi)^{2/3}}.
\end{equation}
Similarly, we get the approximate chemical potential:
\begin{equation}
\label{muga9}
\mu_g \approx\mu_g^{\rm TF2} + O\left(\beta \gamma_y^{1/3}
\ln \gamma_y\right),
\end{equation}
where
\begin{equation}
\mu_g^{\rm TF2}=\frac{\gamma_y+\gamma_z}{2} +\frac{3^{2/3} (\beta^2
\gamma_y\gamma_z)^{1/3}}{2(4\pi)^{2/3}}.
\end{equation}
Specifically, if $\gamma_y =\gamma_z:=\gamma$, then
(\ref{mug91D}), (\ref{ega9}) and (\ref{muga9}) collapse to
\begin{eqnarray}
\label{mu101D}
&&E_g\approx E_g^{\rm TF2}+ O\left(\beta \gamma^{1/3} \ln \gamma\right),
\quad \mu_g\approx \mu_g^{\rm TF2}+
O\left(\beta \gamma^{1/3} \ln \gamma\right), \\
\label{ega13}
&&\mu_{\rm 1D}^{\rm TF}
=\frac{(3\beta\gamma)^{2/3}}{2(4\pi)^{2/3}}, \quad
E_g^{\rm TF2}=\gamma +\frac{3^{5/3}
(\beta\gamma)^{2/3}}{10(4\pi)^{2/3}},\quad
\mu_g^{\rm TF2}=\gamma +\frac{3^{2/3}
(\beta\gamma)^{2/3}}{2(4\pi)^{2/3}}. \qquad
\end{eqnarray}

 To verify (\ref{gsa9}), (\ref{mu101D}) and (\ref{ega13}) numericallly,
   Tables  5 and 6 list the errors
$\|\phi_g^2({\bf x})-(\phi_g^{\rm TF2}({\bf x}))^2\|_{L^1}$ 
and $\|\phi_g^2({\bf x})-(\phi_g^{\rm TF2}({\bf x}))^2\|_{L^\infty}$,
respectively, and
Figure 4 shows the error $\frac{|E_g - E_g^{\rm TF2}|}{E_g}$, for
different $\beta$ and $\gamma$.

\begin{table}[t!]
\begin{center}
\begin{tabular}{lccccc}\hline
$1/\gamma_z$ & $1/12.5$ &$1/25$ &$1/50$   &$1/100$ &$1/400$ \\
\hline

$\beta=25$  &0.2360 &0.1958 &0.1606 &0.1309 &0.1061 \\
rate &  &0.27 &0.29 &0.30 &0.30 \\ \hline

$\beta=50$  &0.3385 &0.2851 &0.2373 &0.1958 &0.1604 \\
rate &  &0.25 &0.26 &0.28 &0.29 \\ \hline

$\beta=100$  &0.4691 &0.4022 &0.3407 &0.2857 &0.2375\\
rate &  &0.22 &0.24 &0.25 &0.27 \\ \hline

$\beta=200$  &0.6212 &0.5440 &0.4706 &0.4026 &0.3408\\
rate &  &0.19 &0.21 &0.23 &0.24 \\ \hline

$\beta=400$  &0.7856 &0.7031 &0.6221 &0.5442 &0.4707\\
rate &  &0.16 &0.18 &0.19 &0.21 \\ \hline

\end{tabular}
\end{center}
\caption{Error analysis of
$\|\phi_g^2-(\phi_g^{\rm TF2}({\bf x}))^2\|_{L^1}$
for the ground state in 3D with a cigar-shaped trap.}
\end{table}

\begin{table}[t!]
\begin{center}
\begin{tabular}{lccccc}\hline
$1/\gamma_z$  & $1/12.5$  & $1/25$ &$1/50$ &$1/100$ &$1/200$   \\
\hline

$\beta=25$  &0.1226 &0.1447 &0.1711 &0.2025 &0.2390 \\

 $\beta=50$  &0.1633 &0.1985 &0.2402 &0.2880 &0.3426 \\

 $\beta=100$  &0.2110 &0.2597 &0.3158 &0.3794 &0.4516 \\
 $\beta=200$  &0.2517 &0.3100 &0.3769 &0.4531 &0.5393 \\
 $\beta=400$  &0.2772 &0.3410 &0.4146 &0.5001 &0.5975 \\
\hline
\end{tabular}
\end{center}
\caption{Error analysis of
$\|\phi_g-\phi^{\rm TF2}\|_{L^\infty}$
for the ground state in 3D with a disk-shaped trap.}
\end{table}

 \begin{figure}[t!]
\centerline{(a)\psfig{figure=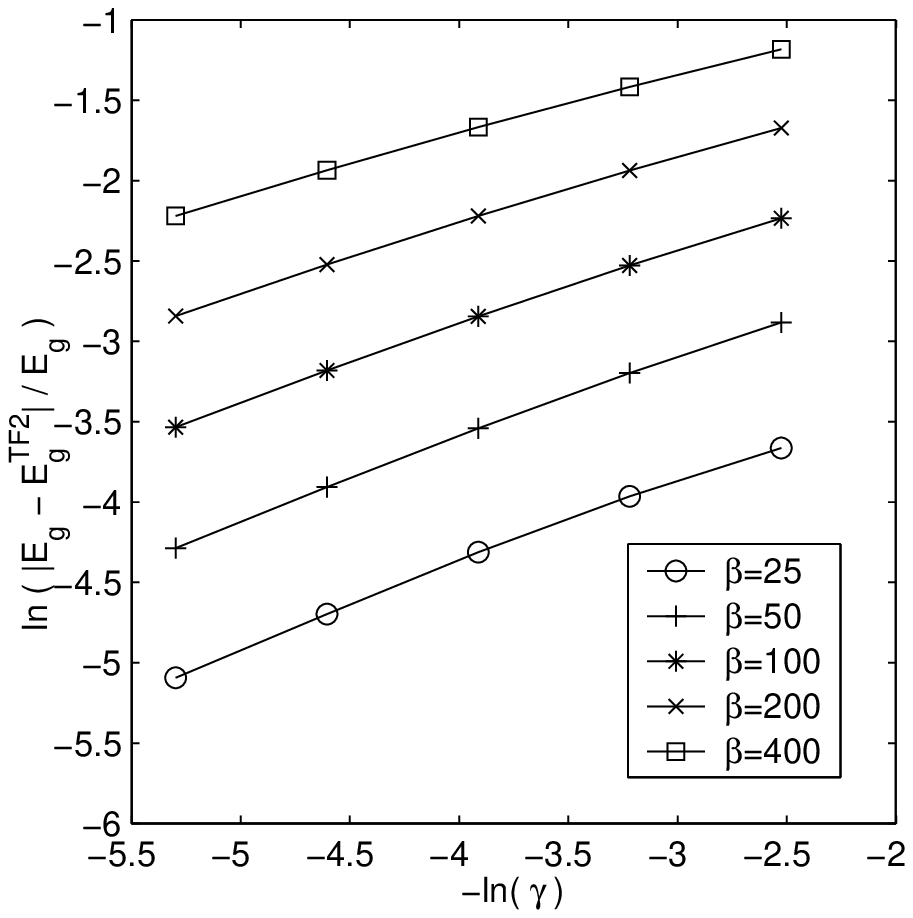,height=7cm,width=6.5cm,angle=0}
\quad (b)\psfig{figure=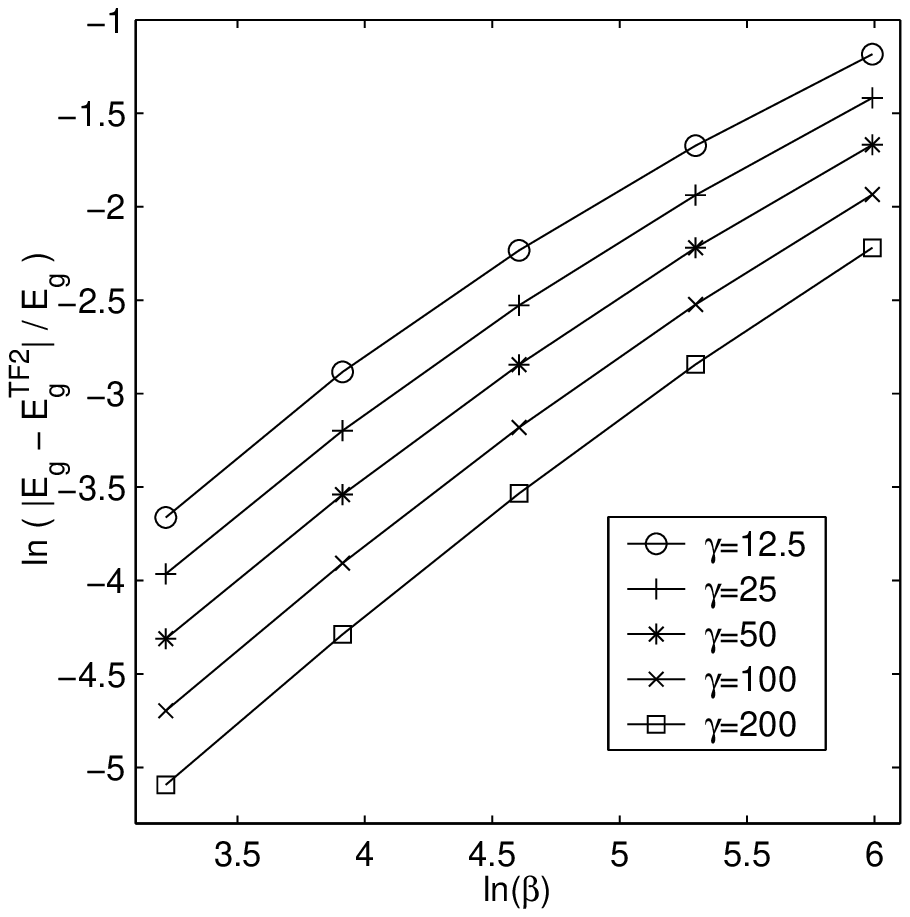,height=7cm,width=6.5cm,angle=0}  }

Figure 4: Convergence rate of $\frac{|E_g - E_g^{\rm TF2}|}{E_g}$
in 3D with a cigar-shaped trap: (a) With respect to $\gamma$; (b)
with respect to $\beta$.
\end{figure}

 From Tabs. 5-6, Fig. 4 and  additional numerical
results in \cite{GeY}, when $\gamma:=\gamma_y=\gamma_z\gg1$,
$\beta_1^a=\beta\gamma/2\pi\gg1$ and $\beta\gamma^{-1}=o(1)$, we
can draw the following conclusion:
\begin{eqnarray*}
&&\|\phi_g-\phi_g^{\rm TF2}\|_{L^2}=
O\left(\frac{\beta^{1/3}\;\ln\gamma}{\gamma^{1/3}}\right), \quad
\|\phi_g^2-(\phi_g^{\rm TF2})^2\|_{L^1}=
O\left(\frac{\beta^{1/3}\;\ln\gamma}{\gamma^{1/3}}\right), \\
&&|E_g-E_g^{\rm TF2}|=
O\left(\beta\; \gamma^{1/3} \ln\gamma\right), \qquad
\frac{|E_g-E_g^{\rm TF2}|}{E_g}=
O\left(\frac{\beta^{1/3}\; \ln\gamma}{\gamma^{2/3}}\right), \\
&&|\mu_g-\mu_g^{\rm TF2}|=
O\left(\beta\; \gamma^{1/3} \ln\gamma\right), \qquad
\frac{|\mu_g-\mu_g^{\rm TF2}|}{\mu_g}=
O\left(\frac{\beta^{1/3}\; \ln\gamma}{\gamma^{2/3}}\right).
\end{eqnarray*}
These results confirm the asymptotic results (\ref{mu101D}),
(\ref{ega13}), (\ref{ega9}) and (\ref{muga9}). Furthermore, 
the additional numerical results in \cite{WW,GeY} indicate that
$\phi_g^{\rm TF2}({\bf x})$ does not converge pointwise to the
ground state
 $\phi_g({\bf x})$ when $\gamma_z\to \infty$ and $\beta>0$ 
(cf. Tab. 6).

\subsection{Choice of initial data for computing ground states
of 3D GPE}

  In order to numerically compute the ground state of the 3D  GPE
(\ref{nleg1}) efficiently, for a given numerical method,
an appropriate choice of the initial data is also very important such
that the iterative number is highly reduced and thus the total
computational cost is greatly saved \cite{WW,WD,MSM}. Based on the
discussion in the previous subsections, here we provide an
appropriate choice for the initial data for different parameter
regimes $\beta$, $\gamma_y$ and $\gamma_z$ in
(\ref{nleg1}):

\begin{itemize}

\item I. When i) $\gamma_y=O(1)$, $\gamma_z=O(1)$ and $\beta=o(1)$
or $O(1)$; ii) $\gamma_y=O(1)$, $\gamma_z\gg1$ and
$\beta_2^a=\beta\sqrt{\gamma_z/2\pi}=o(1)$; or iii)
 $\gamma_y\gg1$, $\gamma_z\gg1$ and
$\beta_1^a=\beta\sqrt{\gamma_y\gamma_z}/2\pi=o(1)$; an appropriate
choice of the initial data is
\[\phi_0(\bx)=\phi_{\rm
ho}(x,y,z)=\frac{(\gamma_y\gamma_z)^{1/4}}{\pi^{3/4}}\;
e^{-\frac{x^2+\gamma_{_y} y^2 +\gamma_{_z} z^2}{2}}, \qquad \bx \in
{\Bbb R}^3.\]

\item  II. When $\gamma_y=O(1)$, $\gamma_z\gg1$,
$\beta_2^a=\beta\sqrt{\gamma_z/2\pi}=O(1)$ and $\beta
\gamma_z^{-3/2}=o(1)$, an appropriate choice of the initial data
is
\[\phi_0(\bx)=\phi_g^{\rm 2D}(x,y)\; \phi_{\rm ho}(z) =
\frac{\gamma_z^{1/4}}{\pi^{1/4}}\; e^{-\frac{\gamma_{_z} z^2}{2}}\;
\phi_g^{\rm 2D}(x,y), \qquad \bx \in {\Bbb R}^3;\] where
$\phi_g^{\rm 2D}(x,y)$ is the ground state of the 2D GPE
(\ref{nleg6}) and can be computed numerically \cite{WW,WD,MSM}.

\item  III. When $\gamma_y=O(1)$, $\gamma_z\gg1$,
$\beta_2^a=\beta\sqrt{\gamma_z/2\pi}\gg1 $ and $\beta
\gamma_z^{-3/2}=o(1)$, an appropriate choice of the initial data
is
\[\phi_0(\bx)=\phi_{\rm 2D}^{\rm TF}(x,y)\; \phi_{\rm ho}(z) =
\frac{\gamma_z^{1/4}}{\pi^{1/4}}\; e^{-\frac{\gamma_{_z} z^2}{2}}\;
\phi_{\rm 2D}^{\rm TF}(x,y), \qquad \bx \in {\Bbb R}^3;\] where
$\phi_{\rm 2D}^{\rm TF}(x,y)$ is given in (\ref{phidsTF}).

\item  IV. When $\gamma_y\gg1$, $\gamma_z\gg1$,
$\beta_1^a=\beta\sqrt{\gamma_y\gamma_z}/2\pi=O(1)$ and $\beta
(\gamma_y\gamma_z)^{-1/2}=o(1)$, an appropriate choice of the
initial data is
\[\phi_0(\bx)=\phi_g^{\rm 1D}(x)\; \phi_{\rm ho}(y,z) =
\frac{(\gamma_y\gamma_z)^{1/4}}{\pi^{1/2}}\;
e^{-\frac{\gamma_{_y}y^2+\gamma_{_z} z^2}{2}}\; \phi_g^{\rm 1D}(x),
\qquad \bx \in {\Bbb R}^3;\] where $\phi_g^{\rm 1D}(x)$ is the
ground state of the 1D GPE (\ref{nleg8}) and can be computed
numerically \cite{WW,WD,MSM}.

\item  V. When $\gamma_y\gg1$, $\gamma_z\gg1$,
$\beta_1^a=\beta\sqrt{\gamma_y\gamma_z}/2\pi\gg1 $ and $\beta
(\gamma_y\gamma_z)^{-1/2}=o(1)$, an appropriate choice of the
initial data is
\[\phi_0(\bx)=\phi_{\rm 1D}^{\rm TF}(x)\; \phi_{\rm ho}(y,z) =
\frac{(\gamma_y\gamma_z)^{1/4}}{\pi^{1/2}}\;
e^{-\frac{\gamma_{_y}y^2+\gamma_{_z} z^2}{2}}\;\phi_{\rm 1D}^{\rm
TF}(x), \qquad \bx \in {\Bbb R}^3;\] where $\phi_{\rm 1D}^{\rm
TF}(x)$ is given in (\ref{phicsTF}).

\item VI. For all other cases, a good choice of the initial data
is
\[\phi_0(\bx)=\phi_g^{\rm TF}(\bx),
\qquad \bx \in {\Bbb R}^3;\] where $\phi_g^{\rm TF}(\bx)$ is given
in (\ref{gsa2}). In fact, in these cases, the 3D GPE (\ref{nleg1})
cannot be reduced to lower dimensions!

\end{itemize}

   Our numerical experiments showed that the initial data in cases
II-V are much better than initial data often  used in the
current literatures.

\section{Dimension reduction for time-dependent GPE}\label{s3}
\setcounter{equation}{0}

  In this section, we will discuss dimension reduction of the
3D  time-dependent GPE (\ref{gpe2}) in certain limiting frequency
regimes and provide convergence rate of the reduction based on our
numerical results.

\subsection{In disk-shaped condensates}

 In the disk-shaped condensates, i.e. for $\omega_x\approx
\omega_y$ and $\omega_z\gg \omega_x$ ($\Longleftrightarrow$
$\gamma_y\approx1$ and $\gamma_z\gg 1$), the 3D GPE (\ref{gpe2})
can be reduced to 2D GPE by assuming that the time evolution does
not cause excitations along the $z$-axis since they have a larger
energy of at least approximately $\hbar \omega_z$ compared to excitations
along the $x$- and $y$-axis with energies of about $\hbar
\omega_x$. Following the physics literature
\cite{PN,BeyondGPE,WW,WDP,Jackson}, for any fixed $\beta\ge0$ and
when $\gamma_z\to\infty$,
 we assume that the condensation wave function along
the $z$-axis is always well described by the ground state wave
function which is well approximated by the harmonic oscillator in
$z$-direction and set \cite{Jackson, BeyondGPE,WW,WDP}
\begin{equation}
\label{d2d} \psi=\psi_2(x,y,t)\phi_3(z), \
\phi_3(z)=\left(\int_{{\Bbb R}^2}\; |\phi_g({\bf x})|^2
\;dxdy\right)^{1/2} \approx \phi_{\rm ho}(z)=\frac{\gamma_z^{1/4}}
{\pi^{1/4}}e^{-\gamma_{_z} z^2/2},
\end{equation}
where $\phi_g(x,y,z)$  defined in (\ref{minp}) is the ground state
solution of the 3D GPE (\ref{gpe2}). Following the procedure used
in \cite{Jackson, BeyondGPE}, the 3D GPE (\ref{gpe2}) can be
reduced to a 2D GPE with ${\bf x}=(x,y)$:
\begin{equation}
\label{gpe2d} i\;\frac{\partial \psi({\bf x},t)}{\partial
t}=-\frac{1}{2}\nabla^2 \psi({\bf x},t)+ V_2(x,y) \psi({\bf x},t) + \beta_2
|\psi({\bf x},t)|^2\psi({\bf x},t),
\end{equation}
where
\begin{equation}
\label{beta2} \beta_2= \beta \int_{-\infty}^\infty
\phi_3^4(z)\,dz\approx \beta_2^a:=\beta
\sqrt{\frac{\gamma_z}{2\pi}}, \quad
V_2(x,y)=\frac{1}{2}\left(x^2+\gamma_y^2 y^2\right).
\end{equation}

To verify (\ref{d2d}) and (\ref{beta2}) numerically, Table 7 lists
the error
  $\|\phi_3(z)-\phi_{\rm ho}(z)\|_{L^2}$,
and  Figure 5 shows the error
$\frac{|\beta_2-\beta_2^a|}{\beta_2}$ vs. $\gamma_z$ and $\beta$,
for different $\beta$ and $\gamma_z$.






\begin{table}[t!]
\begin{center}
\begin{tabular}{lcccc}\hline
$1/\gamma_z$ & $1/25$ &$1/100$ &$1/400$   &$1/1600$ \\
\hline

$\beta=1$  &1.954E-3 &8.619E-4 &3.547E-4 &1.346E-4 \\
rate &  &0.59 &0.64 &0.70 \\ \hline

 $\beta=10$  &1.056E-2 &3.968E-3 &1.446E-3 &5.149E-4 \\
rate &  &0.71 &0.73 &0.74 \\ \hline

 $\beta=100$  &3.709E-2 &1.339E-2 &4.769E-3 &1.689E-3 \\
rate &  &0.74 &0.75 &0.75 \\ \hline

 $\beta=1000$  &1.132E-1 &4.216E-2 &1.511E-2 &5.364E-3 \\
rate &  &0.71 &0.74 &0.75 \\ \hline

$\beta=10000$  &2.902E-1 &1.256E-1 &4.707E-2 &1.687E-2 \\
rate &  &0.60 &0.71 &0.74 \\ \hline

\end{tabular}
\end{center}
\caption{Error analysis of $\|\phi_3-\phi_{\rm ho}\|_{L^2}$ for
dimension reduction from 3D to 2D.}
\end{table}

 \begin{figure}[t!]
\centerline{(a)\psfig{figure=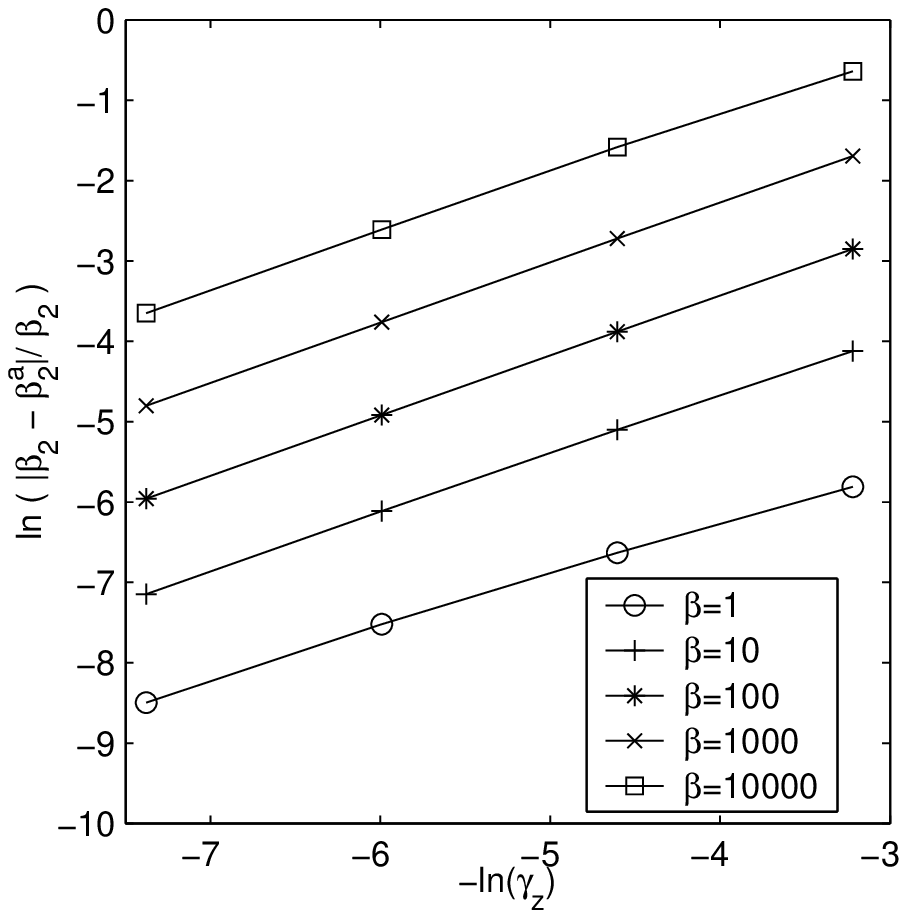,height=7cm,width=6.5cm,angle=0}
\quad (b) \psfig{figure=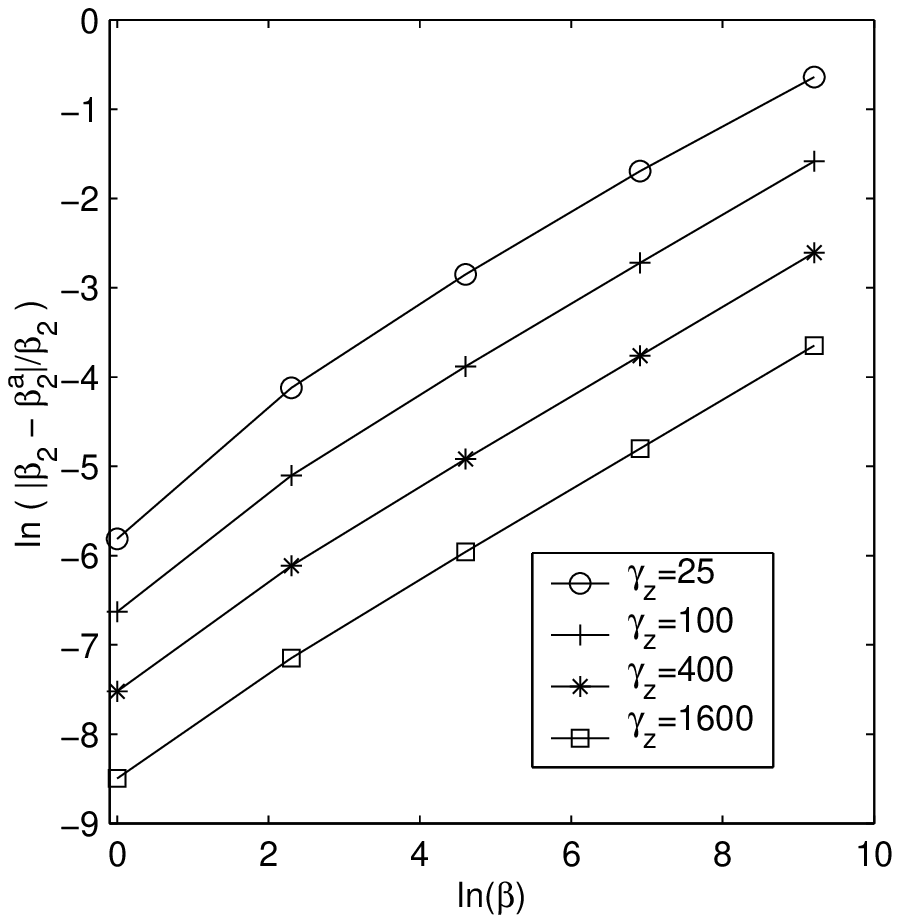,height=7cm,width=6.5cm,angle=0}
}

Figure 5: Convergence rate of
$\frac{|\beta_2-\beta_2^a|}{\beta_2}$ for dimension reduction from
3D to 2D: (a) With respect to $\gamma_z$; (b) with respect to
$\beta$.
\end{figure}

 From Tab. 7, Fig. 5 and  additional numerical
results in \cite{GeY}, for fixed $\beta\ge0$ and when $\gamma_z\to
\infty$,
we can draw the following conclusion:
\begin{eqnarray*}
&&\beta_2 = \beta\sqrt{\frac{\gamma_z}{2\pi}}\left(1+
O\left(\frac{\beta^{1/2}\;
\ln\gamma_z}{\gamma_z^{3/4}}\right)\right), \quad\frac{|\beta_2
-\beta_2^a|}{\beta_2} =
O\left(\frac{\beta^{1/2}\; \ln\gamma_z}{\gamma_z^{3/4}}\right), \\
&&\|\phi_3(z)-\phi_{\rm ho}(z)\|_{L^2}= O\left(\frac{\beta^{1/2}\;
\ln\gamma_z}{\gamma_z^{3/4}}\right), \\
&&\|\phi_3^2(z)-\phi_{\rm ho}^2(z)\|_{L^\infty}=
O\left(\frac{\beta^{1/2}\; \ln\gamma_z}{\gamma_z^{1/2}}\right),
\quad \|\phi_3^2(z)-\phi_{\rm ho}^2(z)\|_{L^1}= O\left(
\frac{\beta^{1/2} \ln\gamma_z}{\gamma_z^{3/4}}\right).
\end{eqnarray*}

In addition, we compare the errors between the solutions of
the 3D GPE (\ref{gpe2}) and its 2D reduction (\ref{gpe2d}) for
different $\gamma_z$. In order to do so, for any given $\gamma_z$,
let $\psi^{\rm 3D}(x,y,z,t)$ be the numerical solution of the 3D
GPE (\ref{gpe2}) with $\gamma_x=\gamma_y=2$,
 $\beta=1$
 and the initial data $\psi(x,y,z,0)=\psi_0(x,y,z)$
chosen as the ground state of (\ref{gpe2}) with
$\gamma_x=\gamma_y=1$, $\beta=1$ \cite{WD,WW}. This 3D dynamics 
 corresponds to a BEC which is in its 
ground state when at   $t=0$ the trap frequencies in
$x$ and $y$ direction are suddently doubled, i.e.
setting $\gamma_{x}= \gamma_{y}=2$. Similarly, let $\psi^{\rm
2D}(x,y,t)$ be the numerical solution of the  2D GPE (\ref{gpe2d})
with $\gamma_x=2$, $\gamma_y=2$, $\beta_2=\beta_2^a:=
\beta\sqrt{\frac{\gamma_z}{2\pi}}$
  and initial data $\psi(x,y,0)=\psi_0(x,y)$
 chosen as the ground state of (\ref{gpe2d}) with
$\gamma_x=\gamma_y=1$,
$\beta_2=\beta_2^a=\beta\sqrt{\frac{\gamma_z}{2\pi}}$
\cite{WD,WW}. Again, this 2D dynamics  corresponds to a BEC 
which is in its 
ground state when at   $t=0$ the trap frequencies in
$x$ and $y$ direction are suddently doubled, i.e.
setting $\gamma_{x}= \gamma_{y}=2$.  In fact,
$\psi^{2D}$ is the solution of the 2D reduction problem. In order
to do the comparison, we introduce
\begin{eqnarray}
\label{dm5} &&\phi_3(z,t)=\left(\int_{{\Bbb R}^2}\left|\psi^{\rm
3D}(x,y,z,t)\right|^2 \;dxdy\right)^{1/2}\approx \phi_{\rm
ho}(z)= \frac{\gamma_z^{1/4}}
{\pi^{1/4}}e^{-\frac{\gamma_{_z} z^2}{2}}, \\
\label{dm77} &&\phi^{\rm 3D}({\bf x},t)\approx \phi^{\rm DS}({\bf
x},t):= \psi^{\rm 2D}(x,y,t)\phi_{\rm ho}(z), \qquad {\bf x}\in
{\Bbb R}^3,
\end{eqnarray}
and the condensate widths
\begin{equation}
\label{cw1} \sigma_\alpha(t)= \int_{{\Bbb R}^3}\alpha^2
\left|\psi^{\rm 3D}({\bf x},t)\right|^2d{\bf x}, \quad
\sigma_\alpha^a(t)= \int_{{\Bbb R}^3}\alpha^2 \left|\psi^{\rm
DS}({\bf x},t) \right|^2d{\bf x},\qquad  \alpha=x,y,z.
\end{equation}

   Figure 6 shows the errors
$\|\psi_3(z,t)-\phi_{\rm ho}(z)\|_{L^\infty}$,
$|\sigma_x-\sigma_x^a|=|\sigma_y-\sigma_y^a|$,
 $\sigma_z-\sigma_z^a=\sigma_z-\frac{1}{4}$ and
 $\left||\psi^{\rm 3D}({\bf 0},t)|^2
-|\psi^{\rm DS}({\bf 0},t)|^2\right|$ for different $\gamma_z$.
Here the numerical solution $\psi^{\rm 3D}$ and $\psi^{\rm 2D}$
are obtained by the time splitting spectral method \cite{WDP,Bao}.

 \begin{figure}[t!]
\centerline{(a)\psfig{figure=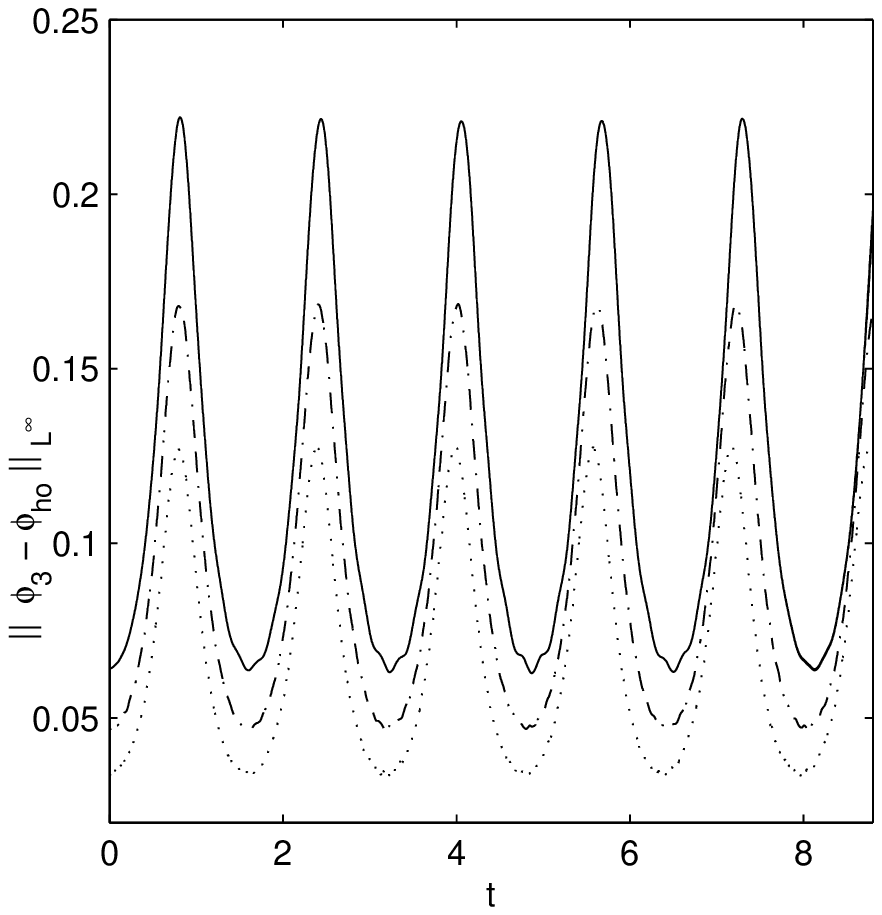,height=5cm,width=12cm,angle=0}}
\centerline{(b)\psfig{figure=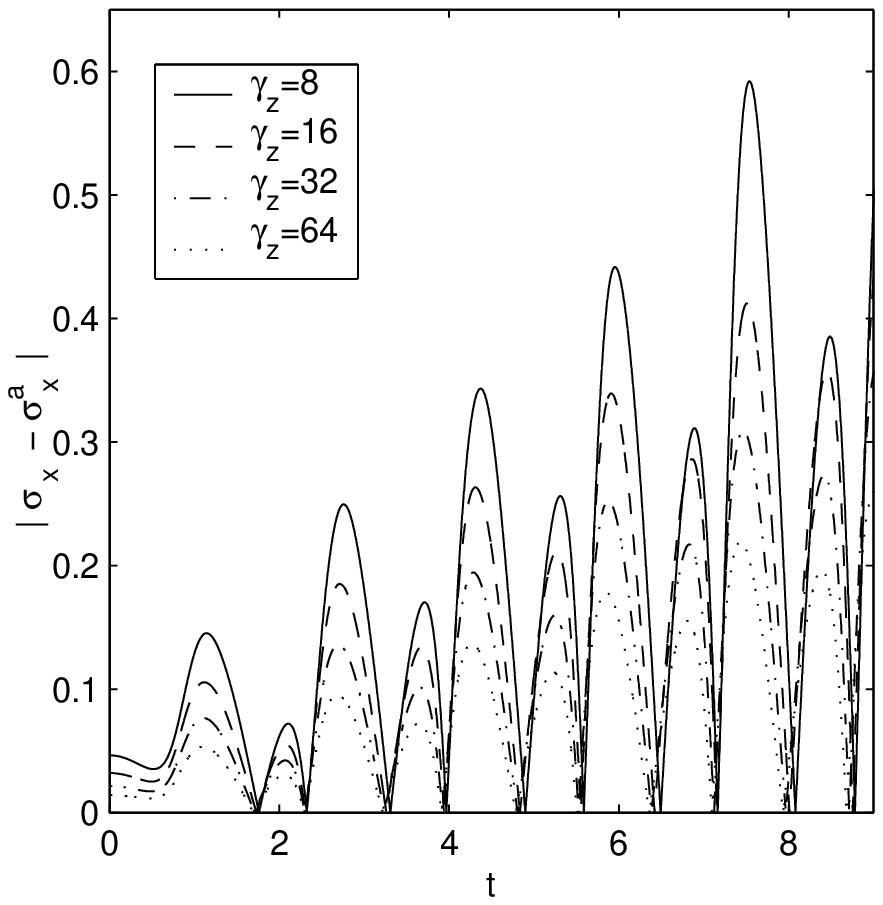,height=5cm,width=12cm,angle=0}}
\centerline{(c)\psfig{figure=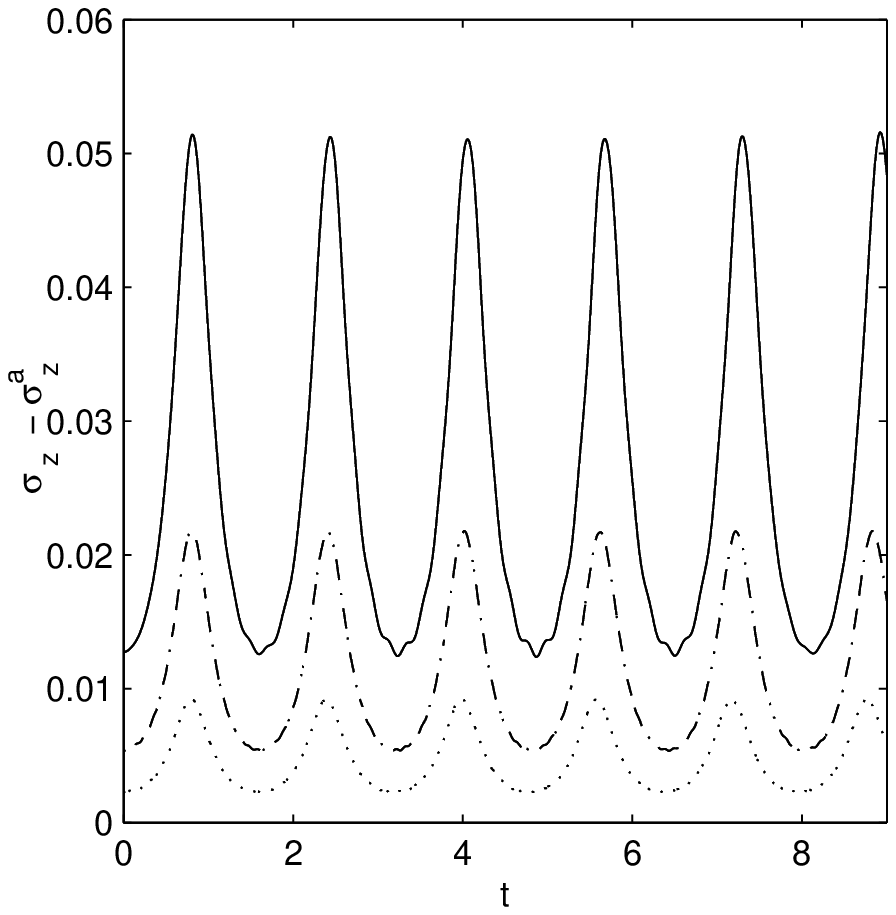,height=5cm,width=12cm,angle=0}
}
 \centerline{(d)\psfig{figure=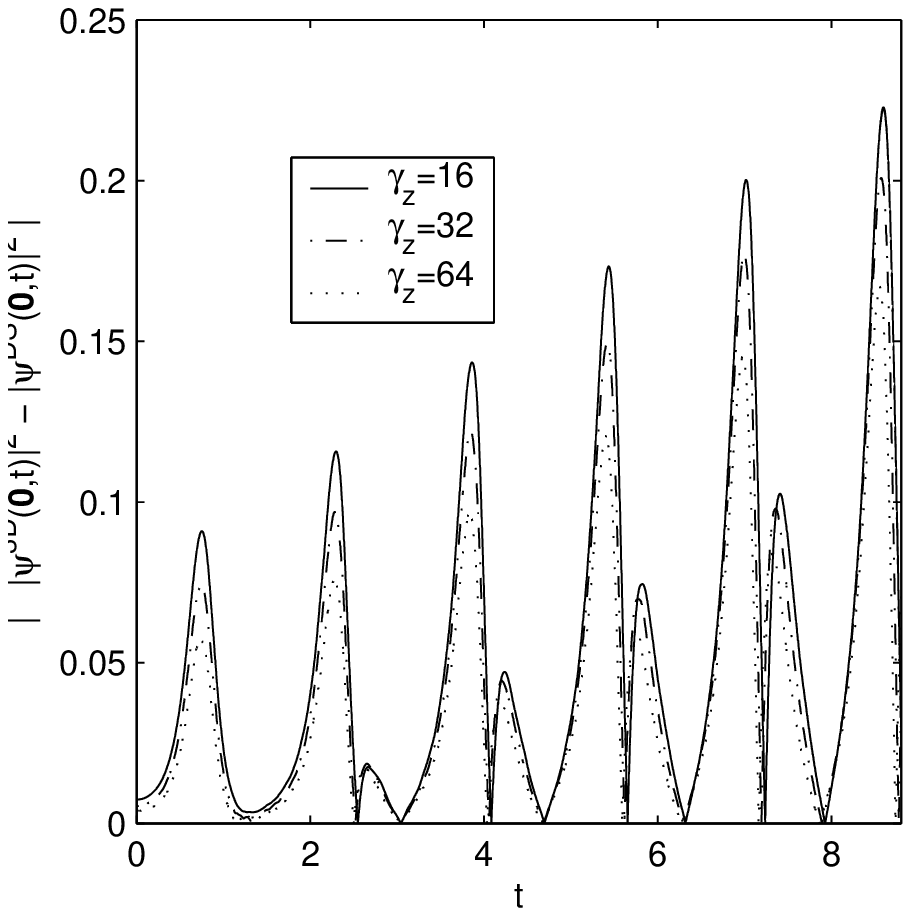,height=5cm,width=12cm,angle=0}  }

Figure 6: Convergence study for dimension reduction from 3D
time-dependent GPE to 2D GPE.

\end{figure}

  From Fig. 6, the dimension reduction from 3D time-dependent
GPE (\ref{gpe2}) to 2D GPE (\ref{gpe2d}) when $\beta=O(1)$ and
$\gamma_z\gg1$ is verified numerically. Furthermore, we have the
following convergence rate:
\begin{eqnarray*}
&&\|\phi_3(z,t)-\phi_{\rm  ho}(z)\|_{L^\infty} =
O\left(\frac{1}{\gamma_z^{3/4}}\right), \qquad
\sigma_x(t)=\sigma_x^a(t) +O\left(\frac{1}{\gamma_z^{3/4}}\right),
\qquad
\gamma_z\gg1,\\
 &&\sigma_z(t)=\frac{1}{4}+O\left(\frac{1}{\gamma_z^{3/4}}\right),  \qquad
|\psi^{\rm 3D}({\bf x},t)|^2= |\psi^{\rm DS}({\bf x},t)|^2
+O\left(\frac{1}{\gamma_z^{1/2}}\right).
\end{eqnarray*}

\subsection{In cigar-shaped condensates}

 Similarly, in the cigar-shaped condensates, i.e.
$\omega_y\gg \omega_x$ and $\omega_z\gg \omega_x$
($\Longleftrightarrow$ $\gamma_y\gg1$ and $\gamma_z\gg 1$), for
any fixed $\beta\ge0$ and when $\gamma_y\to\infty$ and
$\gamma_z\to\infty$, the 3D GPE (\ref{gpe2}) can be reduced to 1D
GPE with ${\bf x}=x$
 \cite{Jackson,PN,WDP,WW,BeyondGPE}:
\begin{equation}
\label{gpe1d} i\; \frac{\partial \psi(x,t)}{\partial
t}=-\frac{1}{2}\partial_{xx} \psi(x,t)+ V_1(x) \psi(x,t)+ \beta_1
|\psi(x,t)|^2\psi(x,t),
\end{equation}
where
\begin{eqnarray}
\label{psi23} &&\beta_1=\beta \int_{{\Bbb
R}^2}|\phi_{23}(y,z)|^4\;dydz\approx \beta_1^a:=\beta
\frac{\sqrt{\gamma_y\gamma_z}}{2\pi}, \qquad
V_1(x)=\frac{x^2}{2}, \\
\label{psi231} &&\phi_{23}(y,z)=\left(\int_{-\infty}^\infty\;
\left|\phi_g(x,y,z)\right|^2 \;dx\right)^{1/2}\approx \phi_{\rm
ho}(y,z):=\frac{(\gamma_y\gamma_z)^{1/4}}{(\pi)^{1/2}}
e^{-\frac{\gamma_{_y} y^2 +\gamma_{_z} z^2}{2}}. \qquad \quad
\end{eqnarray}

Similarly, to verify (\ref{psi23}) and (\ref{psi231}) numerically
with $\gamma:=\gamma_y=\gamma_z$, Table 8 lists the error
  $\frac{\|\phi_{23}(y,z)-\phi_{\rm ho}(y,z)\|_{L^\infty}}
{\|\phi_{23}(y,z)\|_{L^\infty}}$, and Figure 7 shows the errors
$\frac{|\beta_1-\beta_1^a|}{\beta_1}$ vs. $\gamma$ and $\beta$, for
different $\gamma$ and $\beta$.

\begin{table}[t!]
\begin{center}
\begin{tabular}{lccccc}\hline
$1/\gamma$ & $1/12.5$ &$1/25$ &$1/50$   &$1/100$  &$1/400$\\
\hline

$\beta=25$  &0.1727 &0.1412 &0.1145 &0.0924 &0.0743\\
rate &  &0.29 &0.30 &0.31 &0.32 \\ \hline

$\beta=50$  &0.2591 &0.2135 &0.1746 &0.1419 &0.1148\\
rate &  &0.28 &0.29 &0.30 &0.31 \\ \hline

$\beta=100$  &0.3378 &0.3151 &0.2606 &0.2141 &0.1748\\
rate &  &0.26 &0.28 &0.29 &0.29 \\ \hline

$\beta=200$  &0.5334 &0.4517 &0.3791 &0.3156 &0.2608\\
rate &  &0.24 &0.25 &0.26 &0.28 \\ \hline

$\beta=400$  &0.7285 &0.6266 &0.5345 &0.4521 &0.3792\\
rate &  &0.22 &0.23 &0.24 &0.25 \\ \hline

\end{tabular}
\end{center}
\caption{Error analysis of $\frac{\|\phi_{23}(y,z)- \phi_{\rm
ho}(y,z)\|_{L^\infty}}{\|\phi_{23}(y,z)\|_{L^\infty}}$ for
dimension reduction from 3D to 1D.}
\end{table}

 \begin{figure}[t!]
\centerline{(a)\psfig{figure=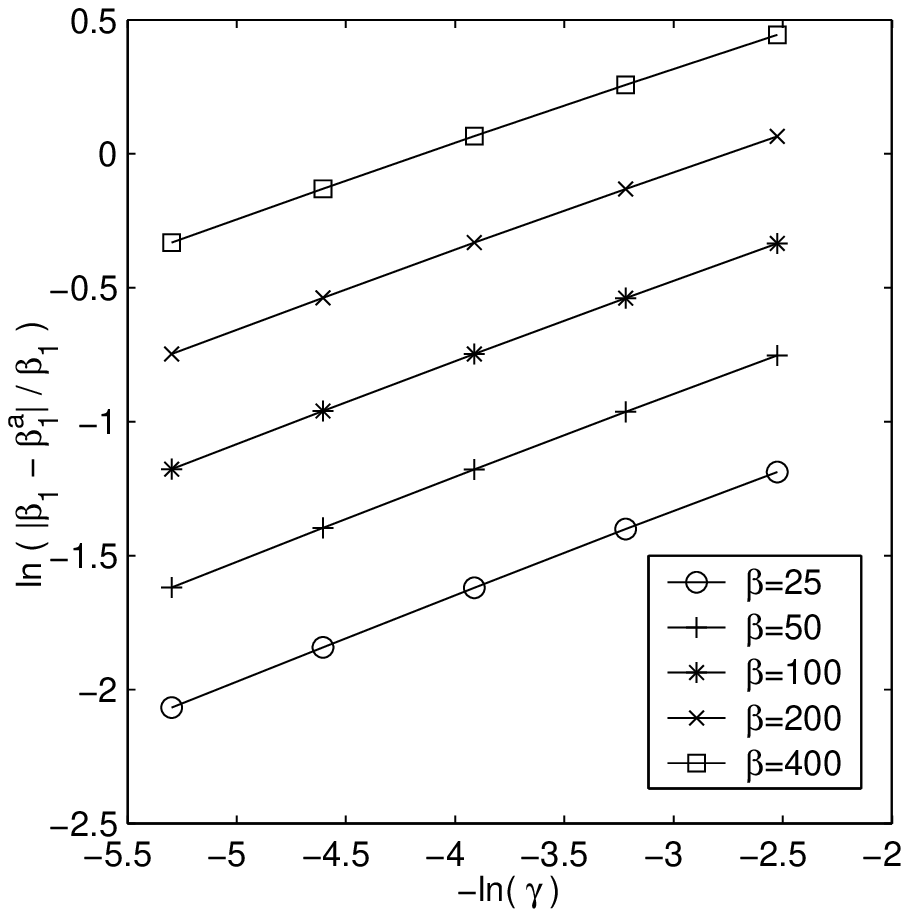,height=7cm,width=6.5cm,angle=0}
\quad (b) \psfig{figure=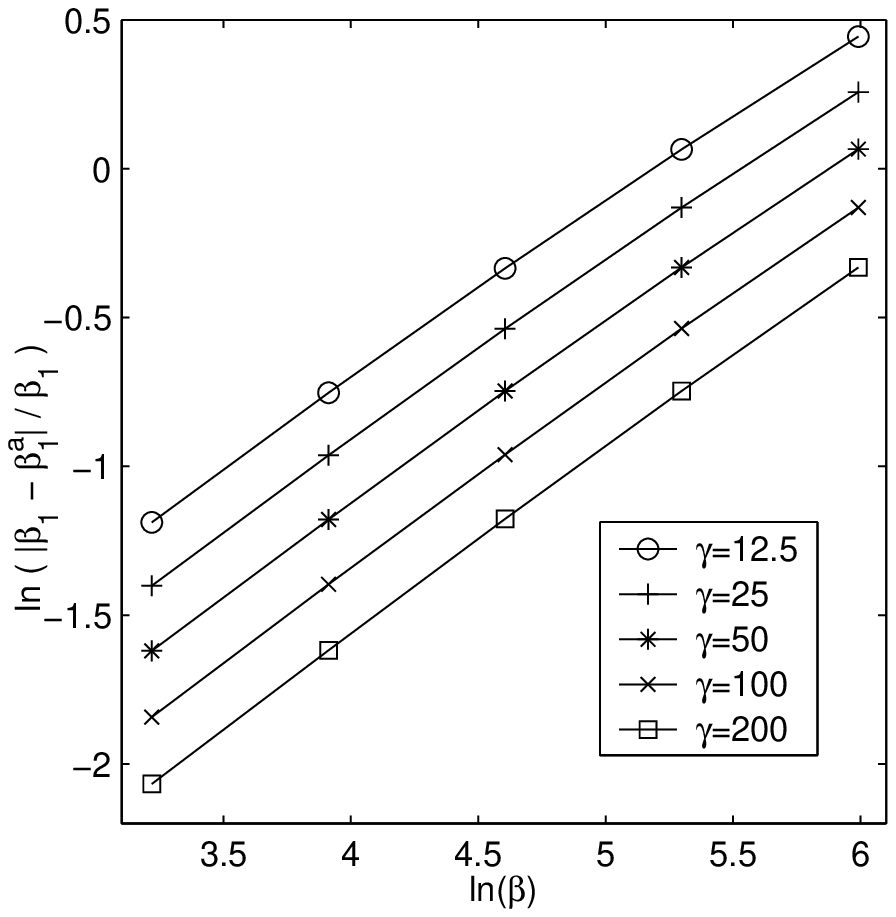,height=7cm,width=6.5cm,angle=0}
}

Figure 7: Convergence rate of
$\frac{|\beta_1-\beta_1^a|}{\beta_1}$ for dimension reduction from
3D to 1D: (a) With respect to $\gamma=\gamma_y=\gamma_z$; (b) with
respect to $\beta$.
\end{figure}

From Tab. 8, Fig. 7 and additional numerical
results in \cite{GeY}, for any fixed  $\beta=O(1)$ and when
$\gamma:=\gamma_y=\gamma_z\to \infty$, we can draw the following
conclusion:
\begin{eqnarray*}
\label{rate1} &&\beta_1 = \beta\frac{\gamma}{2\pi}\left(1+
O\left(\frac{\beta^{1/3} \ln \gamma}{\gamma^{1/3}}\right)\right),
\qquad
\frac{|\beta_1-\beta_1^a|}{\beta_1}=O\left(\frac{\beta^{1/3}\ln
\gamma}
{\gamma^{1/3}}\right), \\
&&\|\phi_{23}(\cdot) - \phi_{\rm ho}(\cdot) \|_{L^\infty}=
O\left(\beta^{1/3}\gamma^{1/3} \ln\gamma \right), \
\frac{\|\phi_{23}(\cdot) - \phi_{\rm ho}(\cdot) \|_{L^\infty}}
{\|\phi_{23}(\cdot)\|_{L^\infty}} =O\left(\frac{\beta^{1/3}\ln
\gamma}
{\gamma^{1/3}}\right), \\
&&\|\phi_{23}^2(\cdot)-\phi_{\rm ho}^2(\cdot)\|_{L^1}=
O\left(\beta^{1/3}\gamma^{1/3} \ln\gamma \right),\
 \frac{\|\phi_{23}^2(\cdot)-\phi_{\rm ho}^2(\cdot)\|_{L^1}}
{\|\phi_{23}^2(\cdot)\|_{L^1}} =
O\left(\frac{\beta^{1/3}\ln\gamma}{\gamma^{1/3}}\right).
\end{eqnarray*}

\subsection{For vortex interaction in 3D}

In this subsection, we numerically study 
dimension reduction for vortex interaction
in 3D disk-shaped condensates. In this case, the assumption 
$\phi_3(z)\approx \phi_{\rm ho}(z)$ in (\ref{d2d}) is no longer valid.
In order to do so, for any given $\gamma_z$, let $\phi_1^{\rm 3D}(x,y,z)$ 
be the central vortex line state with 
winding number $m=1$ of the 3D GPE (\ref{gpe2}) with 
 $\gamma_x=\gamma_y=1$ and $\beta=10$ \cite{WD,WY}; and  
$\psi^{\rm 3D}(x,y,z,t)$ be the 
numerical solution of the 3D GPE (\ref{gpe2}) with $\gamma_x=\gamma_y=1$ and
$\beta=10$ and the initial data
$\psi(x,y,z,0)=\psi_0^{\rm 3D}(x,y,z)$ chosen as 
\[\psi_0^{\rm 3D}(x,y,z)=\frac{\phi_1^{\rm 3D}(x-x_0,y,z)
\phi_1^{\rm 3D}(x+x_0,y,z)}
{\|\phi_1^{\rm 3D}(x-x_0,y,z)\phi_1^{\rm 3D}(x+x_0,y,z)\|_{L^2}},
 \qquad (x,y,z)\in {\Bbb R}^3,\]
where $x_0>0$ is a constant.
This 3D dynamics of BEC corresponds to the interaction of two 
vortex lines in 3D. Formally, when $\gamma_z\gg1$ \cite{WD,WY,WDP}, 
\[\phi_3(z)=\left(\int_{{\Bbb R}^2}|\psi_0^{\rm 3D}(x,y,z)|^2
\;dxdy\right)^{1/2}
\approx \frac{(2\gamma_{_z})^{1/4}}{\pi^{1/4}}\; e^{-\gamma_{_z}z^2}
\neq\phi_{\rm ho}(z).\]
Similarly, let $\phi_1^{\rm 2D}(x,y,z)$ 
be the central vortex state with 
winding number $m=1$ of the 2D GPE (\ref{gpe2d}) with 
 $\gamma_x=\gamma_y=1$ and $\beta_2=\beta\sqrt{\frac{\gamma_z}{2\pi}}$ 
\cite{WD,WY}; and  $\psi^{\rm 2D}(x,y,t)$ be the 
numerical solution of the 2D GPE (\ref{gpe2d}) with $\gamma_x=\gamma_y=1$ and
$\beta_2=\beta\sqrt{\frac{\gamma_z}{2\pi}}$ and the initial data
$\psi(x,y,0)=\psi_0^{\rm 2D}(x,y)$ chosen as 
\[\psi_0^{\rm 2D}(x,y)=\frac{\phi_1^{\rm 2D}(x-x_0,y)\phi_1^{\rm 2D}
(x+x_0,y)}{\|\phi_1^{\rm 2D}(x-x_0,y)\phi_1^{\rm 2D}(x+x_0,y)\|_{L^2}}, 
\qquad (x,y)\in {\Bbb R}^2.\]
Again, this 2D dynamics of BEC corresponds to the interaction of two 
vortices in 2D. Let ${\bf x}_{\rm 2D}(t)=(x(t),y(t))^T$ be the 
location at time $t$ of center of the vortex  initially located at $(x_0,0)$
in the 2D dynamics, and ${\bf x}_{\rm 3D}(t)=(x(t),y(t))^T$ be the 
location in the $xy$-plane at time $t$ of center of the vortex  
initially located at $(x_0,0,0)$ in the 3D dynamics, i.e. center 
of the vortex from the wave function $\psi^{\rm 3D}(x,y,0,t)$.
Figure 8 shows the trajectory of ${\bf x}_{\rm 2D}(t)$ and 
${\bf x}_{\rm 3D}(t)$ with $x_0=1$ for different $\gamma_z$.

 \begin{figure}[t!]
\centerline{(a)\psfig{figure=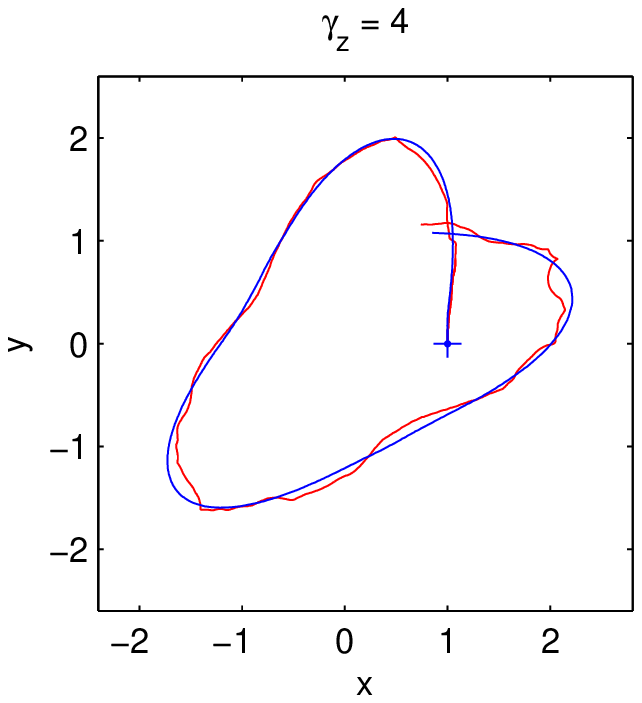,height=7cm,width=6.5cm,angle=0}
\quad (b) \psfig{figure=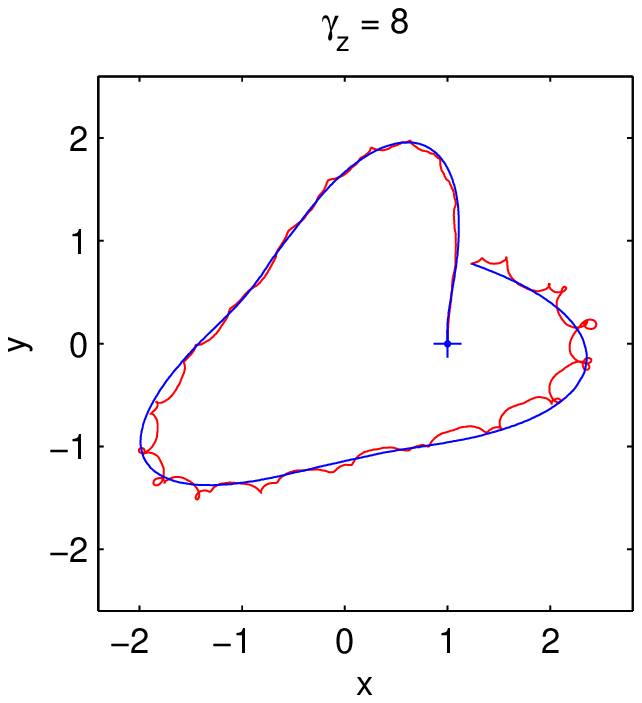,height=7cm,width=6.5cm,angle=0}}
\centerline{(c)\psfig{figure=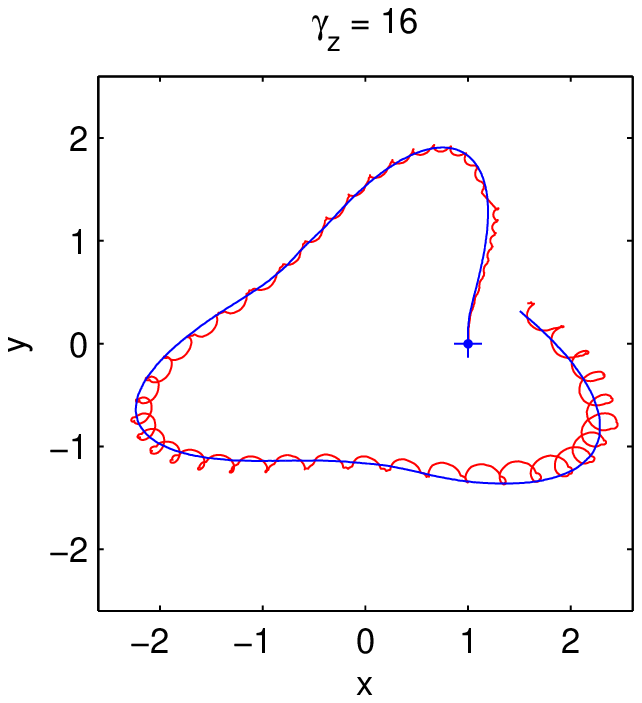,height=7cm,width=6.5cm,angle=0}
\quad (d) \psfig{figure=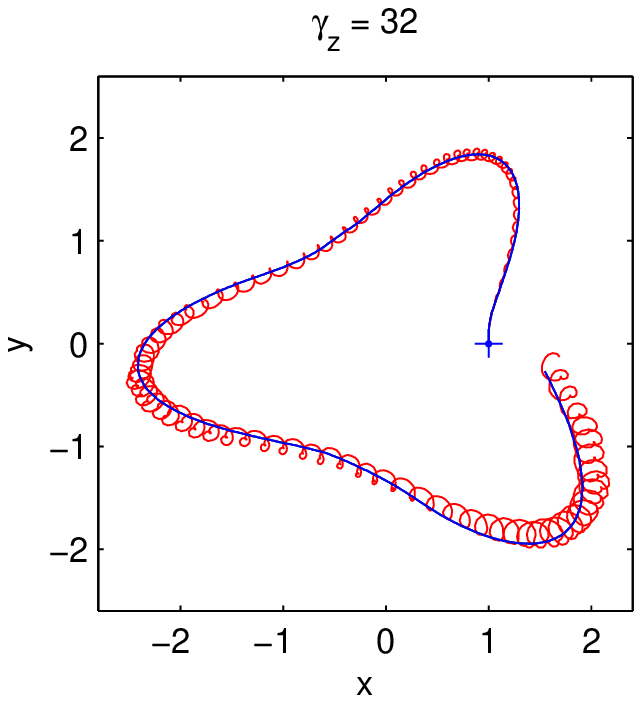,height=7cm,width=6.5cm,angle=0}}

Figure 8: Trajectory of the vortex centers in 2D and 3D dynamics.
Blue line (or flat curve): ${\bf x}_{\rm 2D}(t)$; and
 red line (or oscillatory curve): ${\bf x}_{\rm 3D}(t)$.

\end{figure}

From Fig. 8, for  fixed  $\gamma_z$, the trajectories of the vortex centers
in 2D and 3D dynamics agree qualitatively (cf. Fig. 8).  But
for fixed $\beta=O(1)$, when $\gamma_z\to \infty$, 
the larger is $\gamma_z$, the larger is the error. 
This is because  for larger $\gamma_z$ the condensate becomes
flat and thus the vortex line dynamics in 3D bent more frequently
which causes the oscillatory nature in the trajectory.

  For dimension reduction of the GPE with general initial data, 
we refer to \cite{BaoM,Ben,Sala}.

\section{Conclusion}\label{sc}
\setcounter{equation}{0}

   Dimension reduction of the three-dimensional (3D)
Gross-Pitaevskii equation (GPE) for Bose-Einstein condensation
under different limiting parameter regimes was studied
asymptotically and numerically. For ground state of the 3D GPE, we
found the convergence rate of the reduction and provided the
approximate energy and chemical potential in both disk-shaped
condensation and cigar-shaped condensation. Our extensive
numerical results confirmed the reduction and the convergence rate.
In addition, we identified the parameter regimes in which the
reduction is invalid. For dynamics of the 3D GPE, our numerical
results confirmed the reduction and provided convergence rates in
certain limiting parameter regime.

\bigskip
\bigskip

\setcounter{section}{1}
\renewcommand{\theequation}{\Alph{section}.\arabic{equation}}
\setcounter{equation}{0}
\begin{center} {\bf Appendix:  Energy and chemical potential
approximations for the ground states}
\end{center}

 The 3D time-independent GPE (\ref{nleg1}), 2D GPE (\ref{nleg6}) and
1D GPE (\ref{nleg8}) can be written in a unified way
\cite{BeyondGPE,WDP,WW}
\begin{equation}
\label{nleg1g} \mu\;\phi({{\bf x}}) =  -\frac{1}{2}\nabla^2
\phi({{\bf x}})+V_d({{\bf x}})\phi({{\bf x}}) +\beta_d| \phi({{\bf
x}})|^2\phi({{\bf x}}), \qquad {\bf x}\in{\Bbb R}^d,
\end{equation}
under  the normalization condition
\begin{equation}
\label{gpenorm} \|\phi\|^2:=\int_{{\Bbb R}^d }|\phi({{\bf
x}})|^2d{{\bf x}}=1;
\end{equation}
where $\beta_3=\beta$, $\beta_2=\beta_2^a$, $\beta_1=\beta_1^a$
and $V_3(\bx)=V(\bx)$. The energy functional is defined as
\begin{equation}\label{energy}
E(\phi)=\int_{{\Bbb R}^3}\left[\frac{1}{2}|\nabla\phi({{\bf
x}})|^2+ V({{\bf x}})|\phi({{\bf
x}})|^2+\frac{\beta}{2}|\phi({{\bf x}})|^4 \right]d{{\bf x}}.
\end{equation}

\bigskip

{\bf A1. Thomas-Fermi (TF) approximate energy and chemical
potential.} When $\beta_d\gg1$, and $\gamma_y=O(1)$ with $d=2$ or
 $\gamma_y=O(1)$ and $\gamma_z=O(1)$ with $d=3$ in (\ref{nleg1g}),
we can ignore the kinetic term  and derive the TF approximation:
\begin{equation}
\label{boxTF} \mu_g^{\rm TF}\; \phi_g^{\rm  TF}({\bf x})=
V_d({\bf x}) \phi_g^{\rm TF}({\bf x}) + \beta_d|\phi_g^{\rm
TF}({\bf x})|^2 \phi_g^{\rm TF}({\bf x}), \qquad {\bf x}\in {\Bbb
R}^d.
\end{equation}
Solving (\ref{boxTF}), we obtain the TF approximation for the
ground state:
\begin{equation}
\label{tf3} \phi_g^{\rm TF}({\bf x}) =  \left\{\begin{array}{ll}
\sqrt{\left( \mu_g^{\rm  TF}-V_d({\bf x})\right)/\beta_d},
&\quad V_d({\bf x}) \leq\mu_g^{\rm TF},\\  0, &\quad
\textrm{otherwise}.
\end{array}\right.
\end{equation}
Plugging (\ref{tf3}) into (\ref{gpenorm}) with $\phi=\phi_g^{\rm
TF}$, we obtain \cite{WDP, WW}
\begin{equation}
\label{TFcp} \mu_g^{\rm TF}=\frac{1}{2} \left\{\begin{array}{ll}
\left(\frac{3\beta_1}{2}\right)^{2/3}, &d=1,\\
\left(\frac{8\beta_2 \gamma_y}{2\pi}\right)^{1/2}, &d=2,\\
\left(\frac{15 \beta \gamma_y \gamma_z}{4\pi}\right)^{2/5}, &d=3.\\
\end{array}\right.
\end{equation}
Since $\phi_g^{\rm  TF}({\bf x})$ is not differentiable at
$V_d({\bf x})= \mu_g^{\rm TF}$, as  observed in \cite{WW,WDP,WY},
$E(\phi_g^{\rm TF}) = \infty$, thus one cannot use the definition
(\ref{energy}) to define the energy
 of the TF approximation (\ref{tf3}). Therefore, 
  noticing (\ref{cp1}) and (\ref{ke1}),
as  observed in \cite{WY}, here we used the  way to calculate it:
\begin{equation}
\label{Egtfra} E_g^{\rm TF}\approx  E_g=E(\phi_g)
=\mu(\phi_g)-E_{\rm int}(\phi_g) \approx  \mu_g^{\rm TF} - E_{{\rm
int}}(\phi_g^{\rm TF}) =\frac{d+2}{d+4} \mu_g^{\rm TF}.
\end{equation}

\bigskip

{\bf A2. First-order approximate energy and chemical potential.}

When $\gamma_y=1$ with $d=2$ or $\gamma_y=\gamma_z=1$ with $d=3$
in (\ref{nleg1}), the ground state of the nonlinear eigenvalue
problem (\ref{nleg1g}) is symmetric, i.e. $\phi_g({\bf x}) =
\phi(r)$ with $r=|{\bf x}|$, and satisfies:
\begin{equation}
\label{neg2} -\frac{1}{2r^{d-1}}
\frac{d}{dr}\left(r^{d-1}\frac{d\phi(r)}{dr}\right)
+\left(V_d(r)-\mu\right)\phi(r) +\beta_d \phi^3(r)=0, \quad
0<r<\infty,
\end{equation}
where $V_d(r) = r^2/2$. Following the method used in \cite{LS} for
$d=3$, we choose $R=\sqrt{2\mu_g^{\rm TF}}$ such that
$V_d(R)=\frac{R^2}{2} =\mu_g^{\rm TF}$. When $|r-R|\ll R$, we have
\begin{equation}
\label{pap1} V_d(r)-\mu \approx V_d(r)-\mu_g^{\rm TF} =
\frac{r^2}{2}-\frac{R^2}{2}= (r-R) \frac{r+R}{2} \approx (r-R)R.
\end{equation}
Noticing (\ref{pap1}) and dropping the first order term
$\frac{d-1}{2r}\frac{d\phi(r)}{dr}$ in (\ref{neg2}), we obtain
\begin{equation}
\label{neg3} -\frac{1}{2} \frac{d^2\phi(r)}{dr^2} +(r-R)R\phi(r)
+\beta_d \phi^3(r)=0, \quad 0<r<\infty.
\end{equation}
Introducing a change of variables, $s=(r-R)/l$ and $\phi(r) = \alpha
\tilde{\phi}(s)$ with $2Rl^3=1$ and $2\beta_d \alpha^2 l^2=1$, we
can reduce (\ref{neg3})  to
\begin{equation}
\label{neg4} \tilde{\phi}^{\prime\prime}(s) -(s+\tilde{\phi}^2(s))
\tilde{\phi}(s)=0, \qquad -\infty<s<\infty.
\end{equation}
For the solution of (\ref{neg4}), as $s\to +\infty$,
$\tilde{\phi}\to0$, dropping $\tilde{\phi}^3$ term in (\ref{neg4}),
we have
\begin{equation}
\label{neg5} \tilde{\phi}^{\prime\prime}(s) -s \tilde{\phi}(s)=0.
\end{equation}
Thus we have the asymptotics for the solution \cite{LS}:
\begin{equation}
\label{tphip} \tilde{\phi}(s\to +\infty) \sim \frac{A}{2s^{1/4}}
e^{-\frac{2}{3}s^{2/3}}, \qquad  A\approx 0.794.
\end{equation}
On the other hand,  as $s\to -\infty$, dropping
$\tilde{\phi}^{\prime\prime}(s)$ in (\ref{neg4}), we have
\begin{equation}
\label{neg6} s+ \tilde{\phi}^2(s)=0.
\end{equation}
Thus we get
\begin{equation}
\label{tphim} \tilde{\phi}(s\to -\infty) \sim \sqrt{-s}.
\end{equation}

Choosing $\varepsilon$ such that $l\ll \varepsilon\ll R$, using
$\phi_g \approx \phi_g^{\rm TF}$ for $r\in[0, R-\varepsilon]$ and
$\phi_g \approx \alpha \tilde{\phi}((r-R)/l)$ for $r\in
[R-\varepsilon, \infty)$, plugging (\ref{tf3}), (\ref{tphim}) and
(\ref{TFcp})  with $\gamma_y=\gamma_z=1$ into (\ref{ke1}),
 we get the approximate
kinetic energy of the ground state
\begin{eqnarray}
\label{kenergy} E_{\rm kin}(\phi_g)&=&\frac{1}{2}\int_{{\Bbb R}^d}
|\nabla \phi_g|^2\; d{\bf x} =\frac{C_d}{2} \int_0^\infty
(\phi_g^\prime(r))^2\;
r^{d-1}\; dr\nonumber \\
&=&\frac{C_d}{2}\left[\int_0^{R-\varepsilon}
(\phi_g^\prime(r))^2\; r^{d-1}\; dr +\int_{R-\varepsilon}^\infty
(\phi_g^\prime(r))^2\;
r^{d-1}\; dr\right] \nonumber \\
&\approx&\frac{C_d}{2}\left[\int_0^{R-\varepsilon}
\left(\frac{d\phi_g^{\rm TF}(r)}{dr}\right)^2 \; r^{d-1}\; dr
+\int_{-\varepsilon/l}^\infty \frac{\alpha^2}{l^2}
\left|\tilde{\phi}^\prime(s)\right|^2 (ls +R)^{d-1}l\; ds\right]
\nonumber\\
&=&\frac{C_d}{2}\left[\frac{1}{2\beta_d}\int_0^{R-\varepsilon}
\frac{r^{d+1}}{2\mu_g^{\rm TF}-r^2}\; dr+\frac{\alpha^2
R^{d-1}}{l} \int_{-\varepsilon/l}^\infty
\left|\tilde{\phi}^\prime(s)\right|^2
\left(1+\frac{ls}{R}\right)^{d-1}\;ds
\right] \nonumber\\
&\approx&\frac{C_d}{2}\left[\frac{1}{2\beta_d}\int_0^{R-\varepsilon}
\frac{r^{d+1}}{2\mu_g^{\rm TF}-r^2}\; dr+\frac{\alpha^2
R^{d-1}}{l} \int_{-\varepsilon/l}^\infty
\left|\tilde{\phi}^\prime(s)\right|^2\;ds
\right] \nonumber\\
&\approx&\frac{C_d}{2}\left[\frac{R^d}{4\beta_d}
\left(\ln\frac{R}{2\varepsilon}+D_d\right)+ \frac{\alpha^2
R^{d-1}}{l} \int_{-\varepsilon/l}^\infty
\left|\tilde{\phi}^\prime(s)\right|^2
\sqrt{1+s^2}\;d\ln\left(s+\sqrt{1+s^2}\right) \right] \nonumber\\
&\approx&\frac{C_d}{2}\left[\frac{R^d}{4\beta_d}
\left(\ln\frac{R}{2\varepsilon}+D_d\right) +\frac{\alpha^2
R^{d-1}}{4l }\left(\ln\frac{2\varepsilon}{l} +C\right)
\right] \nonumber\\
&\approx&\frac{C_d}{2}\left[\frac{R^d}{4\beta_d}
\left(\ln\frac{R}{2\varepsilon}+D_d\right) +\frac{ R^d}{4\beta_d
}\left(\ln\left(2\varepsilon (2R)^{1/3}\right)
+C\right) \right] \nonumber\\
&=&\frac{C_d
R^d}{8\beta_d}\left[\ln\left(2^{1/3}R^{4/3}\right)+D_d+C\right]
=\frac{\tilde{C}_d}{\beta_d^{2/(d+2)}}\left(\ln \beta_d
+G_d\right),
 \end{eqnarray}
where
\begin{eqnarray*}
C&=&-4\int_{-\varepsilon/l}^\infty \ln\left(\sqrt{1+s^2}+s\right)
\frac{d}{ds}\left[\sqrt{1+s^2}(\tilde{\phi}^\prime(s))^2\right]\;ds \\
&\approx&-4\int_{-\infty}^\infty \ln\left(\sqrt{1+s^2}+s\right)
\frac{d}{ds}\left[\sqrt{1+s^2}(\tilde{\phi}^\prime(s))^2\right]\;ds
\approx 0.706, \\
R&=&\sqrt{2\mu_g^{\rm TF}} = \left[\frac{((d+1)^2-1)\beta_d}{C_d}
\right]^{1/(d+2)}, \\
\tilde{C}_d&=&\frac{C_d^{2/(d+2)}
((d+1)^2-1)^{d/(d+2)}}{6(d+2)}, \qquad d=1,2,3,\\
G_d&=&\ln \frac{(d+1)^2-1}{C_d} +\frac{d+2}{4} \left(\ln2 + 3D_d
+3C\right),
\end{eqnarray*}

\[ C_d=\left\{\begin{array}{l}
2,\\
2\pi,\\
4\pi,\\
\end{array}\right. \qquad
D_d=\left\{\begin{array}{ll}
\ln4 -2, &\qquad d=1\\
-1, &\qquad d=2 \\
\ln 4-8/3, &\qquad d=3.\\
\end{array}\right.
\]
From (\ref{energy}), (\ref{Egtfra}), (\ref{cp1}), (\ref{ke1}),
(\ref{boxTF}) and (\ref{kenergy}), we can get the first order
approximation for $E_g$ and $\mu_g$ when $\beta_d\gg1$:
\begin{eqnarray}
\label{faenergy} E_g&\approx&E_g^{\rm TF} + E_{\rm
kin}(\phi_g)\approx
 \frac{d+2}{2(d+4)}\left[\frac{((d+1)^2-1)\beta_d}{C_d}
\right]^{2/(d+2)} + \frac{\tilde{C}_d}{\beta_d^{2/(d+2)}}
\left(\ln \beta_d +G_d\right) \nonumber \\
&=&\frac{d+2}{2(d+4)}\left[\frac{((d+1)^2-1)\beta_d}{C_d}
\right]^{2/(d+2)} +O\left(\frac{\ln
\beta_d}{\beta_d^{2/(d+2)}}\right),
\end{eqnarray}
\begin{eqnarray}
\label{famu1} \mu_g&\approx&\mu_g^{\rm TF} + E_{\rm
kin}(\phi_g)\approx
\frac{1}{2}\left[\frac{((d+1)^2-1)\beta_d}{C_d} \right]^{2/(d+2)}
+\frac{\tilde{C}_d}{\beta_d^{2/(d+2)}}
\left(\ln \beta_d +G_d\right) \nonumber \\
&=&\frac{1}{2}\left[\frac{((d+1)^2-1)\beta_d}{C_d}
\right]^{2/(d+2)} +O\left(\frac{\ln
\beta_d}{\beta_d^{2/(d+2)}}\right).
\end{eqnarray}
These asymptotic results were confirmed by the numerical results
in \cite{WW}.

\bigskip

\begin{center}
{\large \bf Acknowledgment}
\end{center}
The first two authors acknowledge support  by the National
University of Singapore  grant No. R-146-000-083-112. P.A.M.
acknowledges support from the EU-funded TMR network `Asymptotic
Methods in kinetic Theory',
 from  his WITTGENSTEIN-AWARD 2000
funded by the Austrian National Science Fund  FWF. W.B. also
acknowledges support from the WITTGENSTEIN-AWARD of P. Markowich
and the hospitality of the Wolfgang Pauli Institute in Vienna for
his extended visit.


\end{document}